\documentclass[12pt]{article}
\usepackage[a4paper,margin=2.5cm]{geometry}
\usepackage{amsmath,amssymb,amsthm}
\usepackage{graphicx}
\usepackage{bm}

\newcommand{\ket}[1]{\left|#1\right\rangle}
\newcommand{\bra}[1]{\left\langle#1\right|}
\newcommand{\braket}[2]{\left\langle#1\middle|#2\right\rangle}
\newcommand{\mel}[3]{\left\langle#1\middle|#2\middle|#3\right\rangle}
\newcommand{\Hil}{\mathcal{H}}
\newcommand{\Aug}{\Psi}
\newcommand{\Res}{R}
\newcommand{\ii}{\mathrm{i}}
\newcommand{\dd}{\mathrm{d}}

\newcommand{\Cplx}{\mathbb{C}}

\newcommand{\id}{1}
\newtheorem{theorem}{Theorem}

\begin{document}

\title{Configuration averaging of X-ray absorption spectra of disordered
systems within the augmented-space full multiple-scattering formalism}

\author{M. Benfatto\thanks{Corresponding author:
\texttt{maurizio.benfatto@lnf.infn.it}}\\
\small INFN, Laboratori Nazionali di Frascati, Via E. Fermi 54,
I-00044 Frascati (RM), Italy}

\date{\today}

\maketitle

\begin{abstract}
We present a formulation of the full multiple-scattering theory of X-ray
absorption spectroscopy (XAS) for disordered systems based on the
augmented-space method of Mookerjee. Both
substitutional (chemical) and thermal (vibrational) disorder are cast, on
the same footing, as exact matrix elements of a \emph{non-random} operator
acting on an enlarged Hilbert space. The configuration-averaged
scattering-path operator is thereby obtained by inverting a non-random
secular matrix, with no expansion in scattering paths, without recourse to the single-site approximation and
without any assumption on the shape of the disorder distribution. This is
what the quantitative analysis of the near-edge region requires: XANES
lies where the multiple-scattering series does not converge, so that a
treatment of disorder tied to a path expansion is unavailable there. The
inversion is carried out by a continued-fraction (Lanczos) recursion,
which accesses the required matrix element without constructing the full
configuration basis or diagonalising the augmented operator. The operator Debye--Waller
factor of the harmonic theory is recovered as the Gaussian special case,
and anharmonic disorder is included with no additional machinery, through
the moments of a non-Gaussian displacement distribution. Where the series
does converge, the construction also settles a question of principle: for
statistically independent site variables, replacing each $t$-matrix by its
configuration average is exact only for
scattering paths in which every site occurs once, and for paths that
revisit a site we obtain the exact second-order correction.
\end{abstract}

\vspace{2ex}

\section{Introduction}
\label{sec:intro}

X-ray absorption spectroscopy (XAS) is an element-selective, local probe
of the geometric and electronic structure around a chosen photoabsorber.
Both its near-edge (XANES) and extended (EXAFS) regions are, from the
real-space multiple-scattering (MST) point of view, a single expression of
Fermi's golden rule evaluated at different energies. That point of view
was established by Schaich~\cite{Schaich1973} and by Lee and
Pendry~\cite{LeePendry1975}, who imported into XAS the machinery developed
for low-energy electron diffraction~\cite{PendryLEED}; it was extended to
the near edge by Durham, Pendry and Hodges~\cite{DurhamPendry1982} and by
the curved-wave treatments developed over the same
years~\cite{Holland1978,MullerSchaich1983,RehrStern1976}.

The two regions differ, however, in one respect that is decisive for what
follows. In the extended region the multiple-scattering series converges
and a few low orders suffice; near the edge it does not, and the
scattering-path operator must be obtained by inverting the secular matrix,
exactly and without any expansion. That full multiple-scattering scheme
grew out of the first-principles cluster calculations of the early
nineteen-eighties~\cite{Kutzler1980,Natoli1980}, where the
extended-continuum treatment of the outer region was introduced, and took
the form of the \textsc{continuum} code~\cite{NatoliBenfatto1986}; the
general-potential formulation from which we start is that of
Refs.~\cite{NatoliPRB90,TysonPRB92}. It is on that scheme that the
quantitative structural analysis of the near-edge region rests:
\textsc{mxan}~\cite{BenfattoMXAN2001,BenfattoMXAN2021} was one of the
first procedures to extract structural parameters from XANES spectra by
fitting them within full multiple scattering, and it remains in use. The
\textsc{feff} codes~\cite{Zabinsky1995,FEFF9,FEFF10}, which came after the
\textsc{continuum} line just described, provide both the path
expansion, which is the economical route wherever the series converges,
and a full multiple-scattering option for the near edge; the theory in the
form in which it is used today is reviewed by Rehr and
Albers~\cite{RehrAlbers2000}.

Real samples are never perfectly
ordered: chemical (substitutional) disorder and thermal (vibrational)
disorder are always present, and the measured spectrum is a configuration
average over the instantaneous atomic arrangements. That the near-edge
region responds to that average, and that reproducing it demands more than
a calculation on the equilibrium geometry, has been established
experimentally at the aluminium $K$ edge, where three different ways of
introducing the vibrations into the cross section were compared against
the measured spectrum~\cite{Manuel2012}.

In XAS the configuration average has been approached along two distinct
lines. The first is analytic: the damping that thermal and structural
disorder produce on the multiple-scattering contributions can be evaluated
in closed form path by path~\cite{Benfatto1989}, and was applied in this
form to the structural disorder of amorphous
silicon~\cite{Filipponi1989b}. Closely related are the cumulant
expansion~\cite{Fornasini2001} and the description of the average in terms
of $n$-body distribution functions~\cite{FilipponiDiCicco1995}. What these
treatments have in common is that the disorder is introduced path by path,
so that they presuppose the very expansion that fails near the edge; they
are accordingly confined to the extended region. The second
line is explicit sampling: a molecular-dynamics trajectory supplies the
instantaneous configurations, the spectrum is computed for each of them,
and the average is taken over the
ensemble~\cite{Roscioni2005,DAngelo2006,Chillemi2016}, an approach
reviewed in Ref.~\cite{BenfattoMXAN2018} and in standard use today: a
recent study of layered systems averages one hundred spectra for Cu and
three hundred for WSe$_2$~\cite{DAcapito2024}. Sampling makes no assumption
on the shape of the distribution and resolves features that a fit with few
structural parameters cannot reach; its cost, however, grows with the
number of configurations required, and the average is performed
numerically, on the spectra, so that it leaves no analytic trace of how
the disorder enters the cross section. The formalism developed here
occupies a middle ground between them: the average is carried out once, at
the level of the operators, before any spectrum is computed, and it is
exact.

The augmented-space formalism of Mookerjee~\cite{Mookerjee1973} provides
the tool. The configuration average of a Green function is mapped onto a
single matrix element of the resolvent of a non-random operator acting on
an enlarged space. The method neither invokes the single-site
approximation nor requires the solution of a self-consistency condition,
and it treats diagonal and off-diagonal disorder on the same footing. It
is worth contrasting it with the closest multiple-scattering alternative,
the alloy analogy model, in which the thermal displacements are themselves
represented as the components of a fictitious alloy and the average is
performed by the single-site CPA~\cite{Ebert2015}. A cluster
generalisation of the same idea, in which the augmented space is combined
with the Korringa--Kohn--Rostoker method and the average is taken by a
self-consistent cluster CPA, is due to Razee \emph{et
al}~\cite{MookerjeePRB90}. All of these share with the present work the
notion of performing the average at the level of the operators, but these
approaches perform the average within an effective-medium approximation
and require a self-consistency condition; the present augmented-space
construction requires neither.
Coupled with a recursion (Lanczos) technique~\cite{Haydock,GrossoPastori}
it has become a standard tool for the electronic structure and the lattice
dynamics of disordered alloys~\cite{ASR}. To our knowledge
it has not been applied to the multiple-scattering \emph{cross section} of
core-level XAS.

Although substitutional disorder is most often discussed in connection
with alloys, it is by no means confined to them. A large class of
systems such as oxides with mixed occupancy, substituted and doped
compounds, solutions and coordination environments, carries a finite
degree of chemical substitution, and typically does so from the second coordination
shell outwards. This is relevant here because, as will be shown in
Sec.~\ref{sec:unified}, the correction that the present treatment supplies
and the standard prescription misses lives on scattering paths that
revisit a site, that is, on the paths that fold back on themselves and
which tend to acquire weight in the outer shells.

In this work we cast the MST expression of the XAS cross section into the
augmented space, and we treat chemical and thermal disorder in one and the
same language. The aim is a theoretical one. The configuration average of
the core-level absorption cross section is performed here exactly within full
multiple-scattering theory, without an effective medium, without a self-consistency
condition and without an assumed form for the disorder distribution; to
our knowledge no complete solution of that problem has been given. The
novelty we would emphasise is not a new expansion but the absence of
one: the disorder is built into the secular matrix before it is inverted,
so that the configuration average becomes available in the same full
multiple-scattering scheme in which XANES spectra are computed, and the
scattering-path operator never has to be expanded in paths. What the paper
offers is the construction itself, the promotion rules, the reference
state, the recursion, and the leading repeated-site correction that the
customary averaged-scatterer prescription misses, in a form meant to be
usable as a specification for a numerical implementation. It
does not offer a material-specific implementation within a production XAS
code, nor a demonstration that the method is competitive on any particular
material: those are separate undertakings, and we say in
Sec.~\ref{sec:concl} what we believe can and cannot be claimed about cost.
Section~\ref{sec:augmented} states the augmented-space
theorem, whose proof is given in Appendix~\ref{app:proof}, and solves the
inverse problem that constructs the disorder operators.
Section~\ref{sec:chem} applies the construction to substitutional
disorder, Sec.~\ref{sec:thermal} recovers the thermal Debye--Waller factor
as a special case, Sec.~\ref{sec:unified} unifies the two and derives the
repeated-site correction, Sec.~\ref{sec:anharm} incorporates anharmonicity
and Sec.~\ref{sec:validation} validates the construction numerically.

\section{The augmented space}
\label{sec:augmented}

\subsection{Configuration average as a matrix element}
\label{sec:theorem}

Let $\Hil$ be a Hilbert space carrying a Hamiltonian $H$, and let
$\{\ket{\theta_n}\}$ be an orthonormal basis of $\Hil$,
$\braket{\theta_n}{\theta_m}=\delta_{nm}$. The resolvent
\begin{equation}
  \Res(z)=\bigl(z\,\id-H\bigr)^{-1},\qquad z\in\Cplx,
  \label{eq:resolvent}
\end{equation}
obeys $\Res(z^{*})=\Res^{\dagger}(z)$ and has all its singularities on the
real axis. We write
\begin{equation}
  G(x,y;E)=\mel{x}{\Res(E)}{y},\qquad \ket{x},\ket{y}\in\{\ket{\theta_n}\}.
  \label{eq:green}
\end{equation}

Suppose now that part of the Hamiltonian is not known deterministically,
but depends on a set of independent stochastic variables
$\{\ell_k\}_{k=1}^{K}$, the variable $\ell_k$ being distributed according
to $p_k(\ell_k)$. The quantity of physical interest is the
configurationally averaged Green function
\begin{equation}
  \overline{G}(x,y;E)=
  \int\!\dd\ell_1\cdots\!\int\!\dd\ell_K\;
  \mel{x}{\Res(\{\ell_k\};E)}{y}\;
  p_1(\ell_1)\cdots p_K(\ell_K).
  \label{eq:avdef}
\end{equation}

For every stochastic variable one may construct an operator $M_k$, acting
on an auxiliary Hilbert space $\phi_k$ with a distinguished normalised
vector $\ket{v_0^{\,k}}$, such that
\begin{equation}
  p_k(\ell_k)=-\frac{1}{\pi}\lim_{z\to\ell_k+\ii 0^{+}}
  \mathrm{Im}\,\mel{v_0^{\,k}}{\bigl(z\,\id-M_k\bigr)^{-1}}{v_0^{\,k}},
  \label{eq:pdos}
\end{equation}
that is, the probability density is realised as the local density of
states of $M_k$ on the state $\ket{v_0^{\,k}}$. The dimension of $\phi_k$
is dictated by $p_k$: it is finite when $\ell_k$ takes a finite number of
values, and infinite for a continuous distribution.

\begin{theorem}[Mookerjee~\cite{Mookerjee1973}]
With the operators $M_k$ defined by \eqref{eq:pdos}, the configuration
average \eqref{eq:avdef} equals a single matrix element of the resolvent
of a non-random operator on the augmented space $\Aug=\Hil\otimes\Phi$,
\begin{equation}
    \overline{G}(x,y;E)=
  \mel{x\otimes\gamma}{\Res\!\bigl(E;\{M_k\}\bigr)}{y\otimes\gamma}
  \label{eq:theorem}
\end{equation}
where
\begin{equation}
  \Phi=\phi_1\otimes\cdots\otimes\phi_K,\qquad
  \ket{\gamma}=\ket{v_0^{\,1}}\otimes\cdots\otimes\ket{v_0^{\,K}},
  \label{eq:gamma}
\end{equation}
and $\Res(E;\{M_k\})$ is obtained from $\Res(E;\{\ell_k\})$ by replacing
each stochastic variable with the corresponding operator acting on its own
factor $\phi_k$.
\end{theorem}

In other words, the stochastic variables are replaced by the operators
$M_k$ and the resolvent is evaluated in the augmented space; the average
is recovered as the matrix element on the ket built with the reference
states $\ket{\gamma}$. The proof is given in Appendix~\ref{app:proof}.

The power of \eqref{eq:theorem} is that the average has become an ordinary,
disorder-free quantum-mechanical matrix element on a larger space. Its
price is the size of $\Phi$: for $N$ sites carrying a binary degree of
freedom, $\dim\Phi=2^{N}$, and the full augmented space has rank
$N_{\Hil}\times 2^{N}$. One therefore never diagonalises the augmented
operator; the matrix element \eqref{eq:theorem} is extracted by a
recursion that needs only matrix--vector products and is controlled by the
truncation level (Appendix~\ref{app:cf}).

\subsection{The inverse problem}
\label{sec:inverse}

The construction requires the solution of an inverse problem: given
$p(u)$, find an operator $M$ and a reference state such that
\begin{equation}
  p(u)=-\frac1\pi\,\mathrm{Im}\,\mel{f_0}{(u\,\id-M)^{-1}}{f_0} .
  \label{eq:inverse}
\end{equation}
$M$ acts on a space $\phi$ whose dimension equals the number of values
that $u$ may assume, the space of the stochastic variable; if $u$
can take infinitely many values, as for a Gaussian, then $\phi$ is
infinite-dimensional. What follows is a general method, valid also for
singular distributions.

We take the reference state to be the first basis vector,
$\ket{f_0}=(1,0,0,\dots)^{\mathsf T}$; this is the state written
$\ket{v_0}$ in Sec.~\ref{sec:theorem}, now expressed in the basis that
tridiagonalises $M$, and the product state $\ket{F}$ used from
Sec.~\ref{sec:promotion} onwards is correspondingly the $\ket{\gamma}$ of
Eq.~\eqref{eq:gamma}. We determine $M$ from its moments
\begin{equation}
  \mu_n=\int u^{\,n}\,p(u)\,\dd u=\mel{f_0}{M^{\,n}}{f_0},
  \qquad \braket{f_0}{f_0}=1 .
  \label{eq:mu_n}
\end{equation}
Writing the local Green function
\begin{equation}
  G_{00}(E)=\mel{f_0}{\frac{1}{E-M}}{f_0}
\end{equation}
and iterating indefinitely the identity
\begin{equation}
  \frac{1}{E-M}=\frac1E+\frac1E\,M\,\frac{1}{E-M}
\end{equation}
one obtains the moment expansion
\begin{equation}
  G_{00}(E)=\sum_{n=0}^{\infty}\frac{\mu_n}{E^{\,n+1}} .
  \label{eq:moments}
\end{equation}
This series is converted into the associated
$J$-fraction~\cite{GrossoPastori}
\begin{equation}
  G_{00}(E)=
  \cfrac{1}{E-a_0-\cfrac{b_1^{2}}{E-a_1-\cfrac{b_2^{2}}{E-a_2-\ddots}}}\,
  \label{eq:jfraction}
\end{equation}
which follows from the corresponding expansion of
\begin{equation}
  F(x)=\cfrac{x}{1-a_0x-\cfrac{b_1^{2}x^{2}}
  {1-a_1x-\cfrac{b_2^{2}x^{2}}{1-a_2x-\ddots}}}
\end{equation}
upon substituting $x=1/E$. The coefficients are given in terms of Hankel
determinants of the moments,
\begin{equation}
  b_n^{2}=\frac{D_n D_{n-2}}{D_{n-1}^{2}}\quad(n\ge1),\qquad
  a_n=\frac{R_n}{D_n}-\frac{R_{n-1}}{D_{n-1}}\quad(n\ge0),\qquad
  D_{-1}=1,\quad R_{-1}=0,
  \label{eq:hankel}
\end{equation}
with
\begin{equation}
  D_n=\begin{vmatrix}
  \mu_0 & \mu_1 & \cdots & \mu_n\\
  \mu_1 & \mu_2 & \cdots & \mu_{n+1}\\
  \vdots & & & \vdots\\
  \mu_n & \mu_{n+1} &\cdots & \mu_{2n}
  \end{vmatrix},
  \qquad
  R_n=\begin{vmatrix}
  \mu_0 & \mu_1 & \cdots & \mu_n\\
  \mu_1 & \mu_2 & \cdots & \mu_{n+1}\\
  \vdots & & & \vdots\\
  \mu_{n-1} & \mu_n &\cdots & \mu_{2n-1}\\
  \mu_{n+1} & \mu_{n+2} &\cdots & \mu_{2n+1}
  \end{vmatrix},
  \label{eq:hankeldef}
\end{equation}
both of order $(n+1)\times(n+1)$; $R_n$ differs from $D_n$ in that the
last row is shifted by one, the row of order $n$ being skipped.

The construction has a transparent reading. Let $\ket{\Psi_\alpha}$ be the
eigenstates of $M$; the local density of states projected on $\ket{f_0}$,
\begin{equation}
  \mu(E)=\sum_\alpha\bigl|\braket{\Psi_\alpha}{f_0}\bigr|^{2}
  \delta(E-E_\alpha)
  =-\frac1\pi\lim_{\varepsilon\to0^{+}}
  \mathrm{Im}\,G_{00}(E+\ii\varepsilon),
\end{equation}
has moments $\mu_n=\int E^{n}\mu(E)\,\dd E$. The distribution $p(u)$ is
therefore a density of states, and $M$ is nothing but its tridiagonal
(Jacobi) representation.

\subsection{Building blocks}
\label{sec:blocks}

\subsubsection{Gaussian distribution}

For
\begin{equation}
  \mu(E)=\frac{1}{\sqrt{2\pi\sigma^{2}}}\,e^{-E^{2}/2\sigma^{2}}
\end{equation}
the moments are
\begin{equation}
  \mu_n=\int E^{n}\mu(E)\,\dd E=
  \begin{cases}
  (n-1)!!\,\sigma^{n}, & n \text{ even},\\[2pt]
  0, & n \text{ odd},
  \end{cases}
\end{equation}
whence
\begin{equation}
  D_0=1,\quad D_1=\sigma^{2},\quad D_2=3\sigma^{6}-\sigma^{6}=2\sigma^{6},
  \quad D_3=12\sigma^{12},\quad\dots
\end{equation}
and, from \eqref{eq:hankel},
\begin{equation}
  b_1^{2}=D_1=\sigma^{2},\qquad
  b_2^{2}=\frac{D_2D_0}{D_1^{2}}=2\sigma^{2},\qquad
  b_3^{2}=\frac{D_3D_1}{D_2^{2}}=3\sigma^{2},\qquad\dots
\end{equation}
that is
\begin{equation}
  b_n=\sigma\sqrt n
  \label{eq:bgauss}
\end{equation}
All the $R_n$ vanish, since every one of these determinants contains a
null row or column: $R_0=\mu_1=0$ and
\begin{equation}
  R_1=\begin{vmatrix}1&0\\ \sigma^{2}&0\end{vmatrix}=0,\qquad
  R_2=\begin{vmatrix}
  1&0&\sigma^{2}\\ 0&\sigma^{2}&0\\ 0&3\sigma^{4}&0
  \end{vmatrix}=0,\qquad\dots
\end{equation}
so that $a_n=0$ for all $n$ and
\begin{equation}
  M=\begin{pmatrix}
  0 & b_1 & 0 & 0 & \cdots\\
  b_1 & 0 & b_2 & 0 & \cdots\\
  0 & b_2 & 0 & b_3 & \cdots\\
  0 & 0 & b_3 & 0 & \ddots\\
  \vdots & & & \ddots & \ddots
  \end{pmatrix},
  \qquad b_n=\sigma\sqrt n .
  \label{eq:mgauss}
\end{equation}
For $\mu(E)=\pi^{-1/2}e^{-E^{2}}$, i.e. $\sigma^{2}=1/2$, this gives
$b_n=\sqrt{n/2}$, as in Ref.~\cite{Mookerjee1973}. The matrix acts on the
disorder space $\phi$, which here is infinite-dimensional because $E$ can
assume infinitely many values distributed as a Gaussian. The state
$\ket{f_0}=(1,0,0,\dots)^{\mathsf T}$ is the reference state in this
space, and $\mel{f_0}{M^{n}}{f_0}=\mu_n$, which is how $M$ was constructed
from the moments of the distribution.

\subsubsection{Delta functions}

This is the distribution relevant to a binary alloy $A_\alpha B_\beta$,
with $\alpha+\beta=1$:
\begin{equation}
  p(E_i)=\alpha\,\delta(E_i-1)+\beta\,\delta(E_i),
  \label{eq:pbin}
\end{equation}
which may also be regarded as the distribution of a continuous variable
$E_i$ subject to
\begin{equation}
  E_i=\begin{cases}1 & \text{if } i=A,\\ 0 & \text{if } i=B.\end{cases}
\end{equation}
Its moments are
\begin{equation}
  \mu_n^{(i)}=\int p(E_i)\,E_i^{\,n}\,\dd E_i=
  \begin{cases}1 & n=0,\\ \alpha & n>0,\end{cases}
\end{equation}
so that
\begin{equation}
  D_0=1,\qquad
  D_1=\begin{vmatrix}1&\alpha\\ \alpha&\alpha\end{vmatrix}
  =\alpha-\alpha^{2}=\alpha\beta,\qquad
  D_2=\begin{vmatrix}1&\alpha&\alpha\\ \alpha&\alpha&\alpha\\
  \alpha&\alpha&\alpha\end{vmatrix}=0,
\end{equation}
whence $b_1=\sqrt{\alpha\beta}$ and $b_n=0$ for $n\ge2$: the matrix $M$
is $2\times2$. For the diagonal coefficients, $R_0=\mu_1=\alpha$ and
\begin{equation}
  R_1=\begin{vmatrix}1&\alpha\\ \alpha&\alpha\end{vmatrix}
  =\alpha-\alpha^{2},
  \qquad
  R_2=\begin{vmatrix}1&\alpha&\alpha\\ \alpha&\alpha&\alpha\\
  \alpha&\alpha&\alpha\end{vmatrix}=0,
\end{equation}
giving
\begin{equation}
  a_0=\frac{R_0}{D_0}=\alpha,\qquad
  a_1=\frac{R_1}{D_1}-\frac{R_0}{D_0}
  =\frac{\alpha-\alpha^{2}}{\alpha-\alpha^{2}}-\alpha=\beta,
\end{equation}
Since the measure is supported on two points, $D_2=0$ and therefore
$b_2^{2}=D_2D_0/D_1^{2}=0$: the Jacobi recursion terminates after the
second basis vector, and no higher coefficient is defined or required.
This is the same structural fact that closes the ternary chain one step
later, $D_3=0$ in Appendix~\ref{app:ternary}. Therefore
\begin{equation}
    M=\begin{pmatrix}\alpha&\sqrt{\alpha\beta}\\[2pt]
  \sqrt{\alpha\beta}&\beta\end{pmatrix}
  \label{eq:mbin}
\end{equation}

It is instructive to run the construction backwards, starting from $M$ and
recovering the distribution; this can be done for any distribution with
finite support. From
\begin{equation}
  \begin{vmatrix}\alpha-\lambda&\sqrt{\alpha\beta}\\[2pt]
  \sqrt{\alpha\beta}&\beta-\lambda\end{vmatrix}=0
  \;\Longrightarrow\;
  (\alpha-\lambda)(\beta-\lambda)-\alpha\beta=\lambda^{2}-\lambda=0,
\end{equation}
the eigenvalues are $\lambda=0$ and $\lambda=1$, as they must be, and the
eigenvectors, expressed on $\ket{f_0}$ and $\ket{f_1}$, are
\begin{equation}
  \ket{\theta_1}=\sqrt\alpha\,\ket{f_0}+\sqrt\beta\,\ket{f_1},
  \qquad
  \ket{\theta_2}=-\sqrt\beta\,\ket{f_0}+\sqrt\alpha\,\ket{f_1},
  \label{eq:eigvec}
\end{equation}
which are readily checked to be orthonormal. Inverting,
\begin{equation}
  \ket{f_0}=\sqrt\alpha\,\ket{\theta_1}-\sqrt\beta\,\ket{\theta_2},
  \qquad
  \ket{f_1}=\sqrt\beta\,\ket{\theta_1}+\sqrt\alpha\,\ket{\theta_2},
\end{equation}
that is $\ket{\theta}=U\ket{f}$ with
\begin{equation}
  U=\begin{pmatrix}\sqrt\alpha&\sqrt\beta\\[2pt]
  -\sqrt\beta&\sqrt\alpha\end{pmatrix},
  \qquad
  U^{-1}=U^{\mathsf T}
  =\begin{pmatrix}\sqrt\alpha&-\sqrt\beta\\[2pt]
  \sqrt\beta&\sqrt\alpha\end{pmatrix}.
\end{equation}
Hence
\begin{equation}
  \mel{f_0}{(z\,\id-M)^{-1}}{f_0}=\frac{\alpha}{z-1}+\frac{\beta}{z},
\end{equation}
and
\begin{equation}
  p(E_i)=-\frac1\pi\lim_{z\to E_i+\ii 0^{+}}\mathrm{Im}
  \left\{\frac{\alpha}{z-1}+\frac{\beta}{z}\right\}
  =\alpha\,\delta(E_i-1)+\beta\,\delta(E_i),
\end{equation}
which is the distribution we started from.

Two properties of \eqref{eq:mbin} will be used repeatedly in what follows.
First, its eigenvalues being $0$ and $1$, $M$ is a projector,
\begin{equation}
  M^{2}=M,
  \label{eq:proj}
\end{equation}
so that $M=\ket{\theta_1}\bra{\theta_1}$. Second,
\begin{equation}
  M\ket{f_0}=\alpha\ket{f_0}+\sqrt{\alpha\beta}\,\ket{f_1},
  \qquad
  \mel{f_0}{M}{f_0}=\alpha,
  \label{eq:Mf0}
\end{equation}
the average of the occupation variable, as it should be.

Nothing in the construction is special to two components. For a ternary
system the same procedure, applied to
$p(E_i)=x\,\delta(E_i-1)+y\,\delta(E_i)+z\,\delta(E_i+1)$, gives
$\mu_0=1$, $\mu_n=x-z$ for odd $n\ge1$ and $\mu_n=x+z$ for even $n$,
whence $D_1=(x+z)-(x-z)^{2}$, $D_2=4xyz$ and $D_3=0$: the disorder
operator is $3\times3$ and the configuration space grows as $3^{N}$. It
is worth noting where this matters. Brute-force averaging becomes
prohibitive one component sooner, whereas in the augmented space the only
change is the dimension of the local block $M^{(i)}$, from two to three: the
structure of the operator, and the cost of the recursion that extracts the
matrix element, are unchanged.

\section{The XAS cross section and substitutional disorder}
\label{sec:chem}

\subsection{The cross section and the choice of the photoabsorber}
\label{sec:crosssec}

Within full multiple-scattering theory the photoabsorption cross section
from a deep core level $\ket{\ell_0 m_0}$ of the atom at the origin
is~\cite{NatoliPRB90,TysonPRB92}
\begin{equation}
  \sigma(\omega)\;\propto\;
  \mathrm{Im}\!\sum_{LL'}D_{L}^{*}\,\tau^{00}_{LL'}\,D_{L'}
  \;=\;\mathrm{Im}\,D^{\dagger}\tau^{00}D,
  \label{eq:sigma}
\end{equation}
where $L=(\ell,m)$, the $D_{L}$ are the dipole matrix elements between the
core state $\ket{\ell_0m_0}$ and the regular solution at the absorbing
site, and $\tau^{00}$ is the site-diagonal block of the scattering-path
operator
\begin{equation}
  \tau=\bigl(T_a^{-1}+G\bigr)^{-1}
      =\bigl(\id+T_aG\bigr)^{-1}T_a
      =\sum_{n\ge0}(-1)^{n}\bigl(T_aG\bigr)^{n}T_a .
  \label{eq:tau}
\end{equation}
Here $T_a$ is the block-diagonal matrix of the atomic $t$-matrices and $G$
the free propagator, both in the normalisation of
Ref.~\cite{TysonPRB92}: the $t$-matrix is dimensionless,
\begin{equation}
  t_\ell=e^{\ii\delta_\ell}\sin\delta_\ell
        =\bigl(\cot\delta_\ell-\ii\bigr)^{-1},
  \label{eq:tmat}
\end{equation}
carrying no factor $-1/k$, and the propagator obeys
$G^{ij}_{00,00}=-e^{\ii kR_{ij}}/(kR_{ij})=-\ii h^{+}_{0}(kR_{ij})$. With
these conventions the sign in front of $G$ is positive, the
multiple-scattering series alternates, and its $n=0$ term reproduces the
atomic cross section, so that the series is normalised to the atomic
value; correspondingly $\sigma$ is proportional to $+\,\mathrm{Im}\,
\tau^{00}$. Throughout this work, and in Fig.~\ref{fig:disorder}, $\chi$
denotes as usual the fractional deviation of the cross section from that
value,
\begin{equation}
  \chi=\frac{\sigma-\sigma_{\mathrm{at}}}{\sigma_{\mathrm{at}}},
  \label{eq:chidef}
\end{equation}
so that the term of order $n$ in \eqref{eq:tau} supplies the contribution
$\chi_n$, and $\chi$ vanishes for an isolated absorber. We work throughout in the extended-continuum regime, in which
the outer-sphere correction to the propagator introduced in
Ref.~\cite{TysonPRB92} is absent; we therefore write $G$ where that work
writes $\tilde G$. Angular-momentum indices are suppressed wherever they
play no role.

The dipole selection rule restricts $L$ to $\ell=\ell_0\pm1$. Whenever the
local symmetry at the absorber, an average over photon polarisations or an
average over magnetic quantum numbers makes the double sum collapse onto
its diagonal, \eqref{eq:sigma} reduces to the more familiar form
$\sum_{L}|D_{L}|^{2}\,\mathrm{Im}\,\tau^{00}_{LL}$; in general, however,
terms with $L\neq L'$ are present, and nothing in what follows requires
that they be absent.

The essential structural feature of \eqref{eq:sigma}, on which everything
that follows depends, is that the cross section is \emph{linear} in
$\tau^{00}$: therefore
\begin{equation}
  \bigl\langle\sigma(\omega)\bigr\rangle\;\propto\;
  \mathrm{Im}\,D^{\dagger}\bigl\langle\tau^{00}\bigr\rangle D
  \label{eq:sigmaav}
\end{equation}
exactly, and the whole problem reduces to the configuration average of
$\tau^{00}$.

At this point a feature specific to XAS must be stated, because it fixes
the statistical ensemble. The absorption edge selects the chemical
species: by tuning the photon energy one decides which atom photoabsorbs,
so that in a binary system $A_xB_y$ measured at the $B$ edge the absorbing
site is occupied by a $B$ atom with probability one. The occupation of the
absorbing site is therefore \emph{not} a random variable, and the ensemble
over which \eqref{eq:sigmaav} is taken is conditioned on it. We accordingly
fix the absorber from the outset: the disorder variables are those of the
$N$ surrounding sites only, and no operator $M^{(0)}$ is attached to the
origin.

This is not a matter of convenience. If the absorbing species were random,
so would be the core-level energy, the regular solution at the origin and
hence the dipole matrix elements $D_{L}$; the average of \eqref{eq:sigma}
would then no longer reduce to that of $\tau^{00}$ alone, with the dipole
factors standing outside it. What would fail is not the augmented-space
construction as such (the core energy, the radial solution and the dipole
elements could in principle be promoted as well) but its reduction to the
single object $\langle\tau^{00}\rangle$, and what one would be computing
is a different and considerably heavier problem. Because the absorber is
fixed, $D_{L}$ is a non-random quantity and may be taken outside the
average.

The two edges of a binary alloy are thus two \emph{distinct} calculations,
$t^{0}=t^{A}$ and $t^{0}=t^{B}$, not two components of a single average.

\subsection{Promotion of the secular matrix}
\label{sec:promotion}

The scattering-path operator is the inverse of the KKR secular matrix
\begin{equation}
  A^{ij}_{LL'}=c^{i}_{LL'}\,\delta_{ij}+G^{ij}_{LL'}\,(1-\delta_{ij}),
  \qquad
  c^{i}=\bigl(T_a^{-1}\bigr)^{ii}=\bigl(t^{i}\bigr)^{-1},
  \label{eq:Amatrix}
\end{equation}
so that $\tau^{00}=[A^{-1}]^{00}$. The disorder enters through the
site-diagonal blocks $c^{i}$, which take one of two values according to
the species occupying site $i$. Introducing the occupation variable
$n_i\in\{0,1\}$, equal to one if site $i$ carries an $A$ atom,
\begin{equation}
  c^{i}=c^{B}+\Delta c\;n_i,
  \qquad \Delta c=c^{A}-c^{B},
  \label{eq:cpromote}
\end{equation}
which is affine in $n_i$. The distribution of $n_i$ is the two-delta
distribution \eqref{eq:pbin} with $\alpha=x$, $\beta=y$, whose disorder
operator is the projector \eqref{eq:mbin},
\begin{equation}
  M^{(i)}=\begin{pmatrix}x&\sqrt{xy}\\[2pt]\sqrt{xy}&y\end{pmatrix},
  \qquad M^{(i)2}=M^{(i)} .
\end{equation}

The theorem of Sec.~\ref{sec:theorem} instructs us to replace $n_i$ by
$M^{(i)}$ acting on the $i$-th factor of the configuration space
$\Phi=\bigotimes_{i=1}^{N}\phi^{(i)}$. The augmented secular matrix is
therefore
\begin{equation}
  \mathcal{A}=
  c^{0}\,\ket{0}\bra{0}\otimes\id_\Phi
  +\sum_{i=1}^{N}\ket{i}\bra{i}\otimes
   \Bigl(c^{B}\id_\Phi+\Delta c\,M^{(i)}\Bigr)
  +\sum_{\substack{i,j=0\\ i\neq j}}^{N}
   G^{ij}\,\ket{i}\bra{j}\otimes\id_\Phi,
  \label{eq:Aaug}
\end{equation}
and the configuration average of the scattering-path operator is a single
matrix element of its inverse,
\begin{equation}
    \bigl\langle\tau^{00}\bigr\rangle
  =\mel{0\,F}{\mathcal{A}^{-1}}{0\,F},
  \qquad
  \ket{F}=\bigotimes_{i=1}^{N}\ket{f_0^{(i)}}
  \label{eq:tauaug}
\end{equation}
with $\ket{f_0^{(i)}}=(1,0)^{\mathsf T}$ the reference state of the $i$-th
factor. The site index runs over the $N+1$ sites of the cluster, the
absorber $i=0$ and its $N$ disordered neighbours, whereas the tensor
product defining $\Phi$ runs over the neighbours alone: the absorber
carries no factor, which is the formal expression of the fact that its
occupation is not a random variable. The states of the augmented space
have the form
$\ket{i}\otimes\ket{f_{j_1}^{(1)}}\otimes\cdots\otimes\ket{f_{j_N}^{(N)}}$,
with each $j_n=0,\dots,\mu-1$ labelling the basis vectors of the $n$-th
configuration factor, whose dimension $\mu$ is the number of components:
the species themselves are the eigenvectors of $M^{(i)}$, Eq.~\eqref{eq:eigvec},
not the basis states $\ket{f_j}$. In the binary case $\mu=2$, $\dim\Phi=2^{N}$, and
the whole augmented space has rank $(N+1)\times2^{N}$. The ternary case,
$\mu=3$, is worked out in Appendix~\ref{app:ternary}. Written out for the
site-diagonal part,
\begin{equation}
  T_B^{-1}=\begin{pmatrix}
   (t^{0})^{-1}&&&\\ &(t_B^{1})^{-1}&&\\ &&\ddots&\\
   &&&(t_B^{N})^{-1}\end{pmatrix},
  \qquad
  \Delta c=\begin{pmatrix}
   0&&&\\ &\Delta c^{1}&&\\ &&\ddots&\\ &&&\Delta c^{N}\end{pmatrix},
  \label{eq:TBDc}
\end{equation}
with $\Delta c^{i}=(t_A^{i})^{-1}-(t_B^{i})^{-1}$. The vanishing first
entry of $\Delta c$ is the absorber: its scatterer is $t^{0}$, fixed by
the edge, and no disorder operator is attached to it.

It is worth writing \eqref{eq:Aaug} out once, on the smallest cluster that
still shows the structure. Take the absorber at the origin, of fixed
species, surrounded by two substitutionally disordered neighbours, and
retain the $s$ channel only, so that each site block is a single number.
The configuration space is $\Phi=\phi^{(1)}\otimes\phi^{(2)}$, of dimension
four; ordering its basis as
\begin{equation}
  \ket{f_0^{(1)}f_0^{(2)}},\quad
  \ket{f_0^{(1)}f_1^{(2)}},\quad
  \ket{f_1^{(1)}f_0^{(2)}},\quad
  \ket{f_1^{(1)}f_1^{(2)}},
\end{equation}
the two local disorder operators are $M^{(1)}=M\otimes\id_2$ and
$M^{(2)}=\id_2\otimes M$, that is
\begin{equation}
  M^{(1)}=\begin{pmatrix}
   x & 0 & \sqrt{xy} & 0\\
   0 & x & 0 & \sqrt{xy}\\
   \sqrt{xy} & 0 & y & 0\\
   0 & \sqrt{xy} & 0 & y
  \end{pmatrix},
  \qquad
  M^{(2)}=\begin{pmatrix}
   x & \sqrt{xy} & 0 & 0\\
   \sqrt{xy} & y & 0 & 0\\
   0 & 0 & x & \sqrt{xy}\\
   0 & 0 & \sqrt{xy} & y
  \end{pmatrix}.
  \label{eq:M12}
\end{equation}
The augmented secular matrix is then the $3\times3$ array of
$4\times4$ blocks
\begin{equation}
  \mathcal{A}=
  \begin{pmatrix}
   c^{0}\,\id_4 & G^{01}\,\id_4 & G^{02}\,\id_4\\[4pt]
   G^{10}\,\id_4 & c^{B}\id_4+\Delta c\,M^{(1)} & G^{12}\,\id_4\\[4pt]
   G^{20}\,\id_4 & G^{21}\,\id_4 & c^{B}\id_4+\Delta c\,M^{(2)}
  \end{pmatrix},
  \label{eq:A12}
\end{equation}
of rank $3\times2^{2}=12$, and the configuration-averaged
scattering-path operator is the single entry
$\langle\tau^{00}\rangle=\mel{0F}{\mathcal{A}^{-1}}{0F}$ with
$\ket{0F}=\ket{0}\otimes(1,0,0,0)^{\mathsf T}$.

Three features of \eqref{eq:A12} are worth noticing, because they persist
for any cluster and any number of channels. The disorder sits entirely in
the site-diagonal blocks; the propagators, which carry all the geometry,
are untouched and simply multiply the identity of the configuration space.
The operator is not random, and no average remains to be taken: the
average has been absorbed into the enlarged matrix. Finally, the two local
operators are \emph{not} interchangeable, as their diagonals show at once:
that of $M^{(1)}$ reads $(x,x,y,y)$ and that of $M^{(2)}$ reads
$(x,y,x,y)$, because each acts on its own factor of the tensor product.
Attaching $M\otimes\id_2$ to both sites would amount to letting a single
occupation variable control two different atoms, and the resulting matrix
element is no longer the configuration average.

A word on the angular basis, which is fixed once and for all here. Angular
momentum indices have been suppressed in \eqref{eq:A12}: each entry is
itself a matrix in $L=(\ell,m)$, and throughout this work these are
expressed in the basis of \emph{real} spherical harmonics, the convention
of Refs.~\cite{NatoliPRB90,TysonPRB92}. The choice is not cosmetic. In
that basis the multiple-scattering matrix is complex symmetric,
$\mathcal{A}^{\mathsf T}=\mathcal{A}$ (the disorder blocks $M^{(i)}$ are
real symmetric and the off-diagonal part is $G\otimes\id_\Phi$), and it is
this property that licenses the three-term recursion by which
$\mathcal{A}^{-1}$ is actually evaluated. In the complex basis the
property fails, and the recursion fails with it, silently. The point is
taken up in Appendix~\ref{app:cf}.

One remark makes the promotion \eqref{eq:cpromote} more useful than it
looks. Since $M^{(i)}$ is a projector, the pair
\begin{equation}
  P_A^{(i)}=M^{(i)},
  \qquad
  P_B^{(i)}=\id_{\phi^{(i)}}-M^{(i)},
  \label{eq:projAB}
\end{equation}
is a complete set of orthogonal projectors on the $i$-th factor of the
configuration space, $P_A^{2}=P_A$, $P_B^{2}=P_B$, $P_AP_B=P_BP_A=0$,
$P_A+P_B=\id$. The promoted scatterer of site $i$ is then a spectral
decomposition,
\begin{equation}
  \mathcal{T}^{i}
  =t_A^{i}\otimes P_A^{(i)}+t_B^{i}\otimes P_B^{(i)},
  \label{eq:Tspectral}
\end{equation}
in which the two angular-momentum blocks live on mutually annihilating
subspaces of $\Phi$. Consequently, for any function $f$ defined on both
blocks,
\begin{equation}
  f\bigl(\mathcal{T}^{i}\bigr)
  =f\bigl(t_A^{i}\bigr)\otimes P_A^{(i)}
  +f\bigl(t_B^{i}\bigr)\otimes P_B^{(i)}
  =f\bigl(t_B^{i}\bigr)\otimes\id
  +\Bigl[f\bigl(t_A^{i}\bigr)-f\bigl(t_B^{i}\bigr)\Bigr]\otimes M^{(i)},
  \label{eq:affine}
\end{equation}
the second form following from $P_B=\id-P_A$. The argument uses only the
orthogonality of the projectors on the configuration factor and therefore
holds unchanged when $t_A^{i}$ and $t_B^{i}$ are matrices in the
angular-momentum indices that do \emph{not} commute with each other, which
is the case of interest; no property of the $t$-matrices themselves is
invoked. In particular
$(\mathcal{T}^{i})^{-1}
 =(t_A^{i})^{-1}\otimes P_A^{(i)}+(t_B^{i})^{-1}\otimes P_B^{(i)}$,
so that promoting the inverse $c^{i}=(t^{i})^{-1}$ and promoting $t^{i}$
itself are simultaneously exact statements, and one may pass freely
between the two.

\subsection{The multiple-scattering series}
\label{sec:MSseries}

For the expansion in scattering paths it is convenient to rewrite the
scattering-path operator without inverting the $t$-matrices. Using
\begin{equation}
  (A+B)^{-1}=\bigl[A(\id+A^{-1}B)\bigr]^{-1}
  =(\id+A^{-1}B)^{-1}A^{-1}
\end{equation}
with $A=T_a^{-1}$ and $B=G$ one obtains
\begin{equation}
  \tau=\bigl(\id+T_aG\bigr)^{-1}T_a,
\end{equation}
which in the augmented space reads
\begin{equation}
  \tau=\bigl(\id\otimes\id_\Phi+\mathcal{K}G\bigr)^{-1}\mathcal{K},
  \qquad
  \mathcal{K}=t^{0}\ket{0}\bra{0}\otimes\id_\Phi
  +\sum_{i=1}^{N}\ket{i}\bra{i}\otimes
   \bigl(t_B^{i}\id_\Phi+\Delta t^{i}M^{(i)}\bigr),
  \label{eq:Kdef}
\end{equation}
with $\Delta t^{i}=t_A^{i}-t_B^{i}$, the promotion of $T_a$ being
legitimate by \eqref{eq:affine}. The multiple-scattering series is then
made of terms of the type
$(\mathcal{K}G)^{n}\mathcal{K}$, and the configuration average of each of
them is an ordinary matrix element on $\ket{F}$.

\paragraph{Single scattering (EXAFS).}
The lowest non-trivial term is $\mathcal{K}G\mathcal{K}G\mathcal{K}$,
whose average
\begin{equation}
  \chi_2\;\sim\;\mel{0F}{\mathcal{K}G\mathcal{K}G\mathcal{K}}{0F}
\end{equation}
describes the path $0\to J\to 0$. With the absorber fixed the first and
last factors carry no disorder operator, and the average reduces to the
single intermediate site,
\begin{equation}
  \chi_2\;\sim\;
  t^{0}\,G^{0J}\,
  \Bigl[t_B^{J}+\bigl(t_A^{J}-t_B^{J}\bigr)
  \mel{F}{M^{(J)}}{F}\Bigr]\,G^{J0}\,t^{0},
\end{equation}
and, by \eqref{eq:Mf0}, $\mel{F}{M^{(J)}}{F}=x$, so that
\begin{equation}
    \chi_2\;\sim\;t^{0}\,G^{0J}\,\bar t^{\,J}\,G^{J0}\,t^{0},
  \qquad
  \bar t=y\,t^{B}+x\,t^{A}
  \label{eq:chi2}
\end{equation}
summed over the neighbours $J$. The prescription of replacing the
$t$-matrix of each scatterer by its configuration average is thus exact at
this order: a single shell of $N$ neighbours behaves as two sub-shells of
populations $Ny$ and $Nx$, which is what the standard analysis of
substitutional alloys assumes.

\paragraph{Third-order scattering.}
The same happens for a path $0\to J\to k\to 0$ with $J\neq k$, in which two
distinct intermediate sites are visited once each. Both promoted
$t$-matrices are expanded, and since the two occupation variables are
independent each contributes its own $\mel{F}{M^{(i)}}{F}=x$; using
$x+y=1$ the four resulting contributions combine into
\begin{equation}
  \begin{split}
  \mel{0F}{\mathcal{K}G\mathcal{K}G\mathcal{K}G\mathcal{K}}{0F}
  &\;\sim\;t^{0}\,G^{0J}
  \Bigl[\,y^{2}\,t_B^{J}G^{Jk}t_B^{k}
  +xy\bigl(t_B^{J}G^{Jk}t_A^{k}+t_A^{J}G^{Jk}t_B^{k}\bigr)\\
  &\hspace{5.2em}
  +x^{2}\,t_A^{J}G^{Jk}t_A^{k}\Bigr]G^{k0}\,t^{0}\\[4pt]
  &\;=\;t^{0}\,G^{0J}\,\bar t^{\,J}\,G^{Jk}\,\bar t^{\,k}\,G^{k0}\,t^{0} .
  \end{split}
  \label{eq:chi3}
\end{equation}
The two cross terms are distinct objects and must be kept apart: the
propagator carries the ordering $J\to k$, so that placing the $A$ atom on
the first site or on the second gives $t_A^{J}G^{Jk}t_B^{k}$ in one case
and $t_B^{J}G^{Jk}t_A^{k}$ in the other, and each occurs once with weight
$xy$. Once they are both retained the bracket factorises into
$\bar t^{\,J}G^{Jk}\bar t^{\,k}$, and again nothing survives beyond the
averaged $t$-matrices.

What both results have in common is that every disordered site along the
path is visited exactly once, so that each promoted $t$-matrix
contributes a single factor $M^{(i)}$ and each such factor is replaced by
its expectation value $x$. Should a path return to a site already visited,
two operators $M^{(i)}$ would meet on the same factor of $\Phi$, and it is
the projector property $M^{2}=M$, not a factorisation into
$\mel{F}{M}{F}^{2}$, that governs their product. This is the mechanism
analysed in Sec.~\ref{sec:unified}, and it is the reason why the averaged
$t$-matrix ceases to be sufficient there.

\section{Thermal disorder}
\label{sec:thermal}

\subsection{Displacement operators}
\label{sec:displ}

We now let the atoms move about their equilibrium positions,
$\vec R_i=\vec R^{\,0}_i+\vec u_i$, and ask how the secular matrix
responds. Only the propagators depend on the positions, and their
dependence can be transferred entirely onto the $t$-matrices. Starting
from the free-electron Green function and the expansion of a plane wave in
spherical waves,
\begin{equation}
  e^{\ii\vec k\cdot\vec r}
  =4\pi\sum_{L}\ii^{\ell}\,j_\ell(kr)\,Y^{*}_{L}(\hat k)\,Y_{L}(\hat r),
\end{equation}
one finds, following Fritzsche~\cite{Fritzsche1994},
\begin{equation}
  G\bigl(\vec R_J+\vec u_J;\vec R_i+\vec u_i\bigr)
  =J(\vec u_J)\;G\bigl(\vec R^{\,0}_{Ji}\bigr)\;J(-\vec u_i),
  \label{eq:Gshift}
\end{equation}
where the displacement matrices are
\begin{equation}
  J_{L'L}(\vec u)=\ii^{\,\ell'-\ell}
  \int\dd\Omega_k\;Y^{*}_{L'}(\hat k)\,Y_{L}(\hat k)\,
  e^{\ii\vec k\cdot\vec u} .
  \label{eq:Jdef}
\end{equation}
They satisfy
\begin{equation}
  J(\vec u)J(-\vec u)=\id,
  \qquad
  J_{L'L}(\vec u)=J_{LL'}(-\vec u)=(-1)^{\ell+\ell'}J_{L'L}(-\vec u).
\end{equation}

Differentiating \eqref{eq:Jdef} with respect to a Cartesian component of
the displacement brings down a factor $\ii k_\alpha$, which can be
re-expressed through $Y_{1\mu}$, giving
\begin{equation}
  \frac{\partial}{\partial u_\alpha}J(\vec u)=\ii k\,M^{\alpha}J(\vec u),
  \label{eq:Jode}
\end{equation}
with generators
\begin{equation}
  M^{z}_{L'L}=\ii^{\,\ell'-\ell}\sqrt{\frac{4\pi}{3}}
  \int\dd\Omega\;Y^{*}_{L'}Y_{L}Y_{10},
  \qquad
  M^{\pm}_{L'L}=\ii^{\,\ell'-\ell}\sqrt{\frac{4\pi}{3}}
  \int\dd\Omega\;Y^{*}_{L'}Y_{L}Y_{1\pm1},
  \label{eq:Mgen}
\end{equation}
and $M^{x}=(M^{-}-M^{+})/\sqrt2$, $M^{y}=\ii(M^{-}+M^{+})/\sqrt2$. These
matrices commute among themselves,
\begin{equation}
  \bigl[M^{\alpha},M^{\beta}\bigr]=0,
  \label{eq:Mcomm}
\end{equation}
and are Hermitian; since they commute, \eqref{eq:Jode} integrates to
\begin{equation}
    J(\vec u)=\exp\Bigl(\ii k\sum_{\alpha}u_\alpha M^{\alpha}\Bigr)
  \label{eq:Jexp}
\end{equation}
For real $k$ the Hermiticity of the generators makes $J(\vec u)$ unitary,
$J(\vec u)^{\dagger}=J(-\vec u)$. When the potential is complex, an
optical potential, or a self-energy accounting for the finite mean free
path, $k$ is complex and unitarity is lost. This is of no consequence
here: what the construction uses is only the invertibility
\begin{equation}
  J^{-1}(\vec u)=J(-\vec u),
  \label{eq:Jinv}
\end{equation}
which follows from \eqref{eq:Jexp} irrespective of whether $k$ is real.
Properties of the generators, and the sense in which \eqref{eq:Mcomm}
holds in a truncated basis, are collected in
Appendix~\ref{app:gener}.

The displaced $t$-matrix follows by moving the displacement operators from
the propagators onto the scatterers,
\begin{equation}
  t(\vec u)=J(-\vec u)\,t^{0}\,J(\vec u)=e^{-A}\,t^{0}\,e^{A},
  \qquad
  A=\ii k\sum_{\alpha}u_\alpha M^{\alpha},
  \label{eq:tsim}
\end{equation}
and, by the Campbell--Baker--Hausdorff theorem,
\begin{equation}
  e^{xA}Be^{-xA}=B+x[A,B]+\frac{x^{2}}{2!}\bigl[A,[A,B]\bigr]+\cdots
  \equiv e^{x[A,\,\cdot\,]}B .
\end{equation}
Equation~\eqref{eq:tsim} is the case $x=-1$, so that, the generators
commuting,
\begin{equation}
  \begin{split}
  t(\vec u)&=\exp\Bigl(-\ii k\sum_{\alpha}u_\alpha
  \bigl[M^{\alpha},\,\cdot\,\bigr]\Bigr)\,t^{0}\\
  &=t^{0}-\ii k\sum_{\alpha}u_\alpha\bigl[M^{\alpha},t^{0}\bigr]
  -\frac{k^{2}}{2!}\sum_{\alpha\beta}u_\alpha u_\beta
  \bigl[M^{\alpha},[M^{\beta},t^{0}]\bigr]+\cdots
  \end{split}
  \label{eq:tu}
\end{equation}
The sign of the exponent is fixed by the order in which the two
displacement matrices flank $t^{0}$ in \eqref{eq:tsim}, and it affects the
odd orders only. The quadratic term is insensitive to it, since it carries
$(\pm\ii)^{2}$, and so therefore is the Gaussian resummation of the next
subsection; the terms proportional to an odd cumulant, and in particular
the third-cumulant contributions of Sec.~\ref{sec:anharm}, do change sign.

\subsection{The Gaussian average and the operator Debye--Waller factor}
\label{sec:DW}

For harmonic vibrations, $\langle\vec u\rangle=0$ and the average of an
exponential is Gaussian~\cite{Maradudin},
\begin{equation}
  \Bigl\langle\exp\Bigl(-\ii\sum_\alpha k_\alpha u_\alpha\Bigr)
  \Bigr\rangle
  =\exp\Bigl(-\tfrac12\sum_{\alpha\beta}
  \langle u_\alpha u_\beta\rangle k_\alpha k_\beta\Bigr).
\end{equation}
Applying the same result to \eqref{eq:tu} and assuming isotropic
vibrations, $\langle u_\alpha u_\beta\rangle=\langle u^{2}\rangle
\delta_{\alpha\beta}\equiv\sigma^{2}\delta_{\alpha\beta}$, one obtains the
averaged $t$-matrix
\begin{equation}
    \bar T=\bigl\langle J(-\vec u)\,t^{0}\,J(\vec u)\bigr\rangle
  =\exp\Bigl(-\tfrac12k^{2}\sigma^{2}\sum_{\alpha}
  \bigl[M^{\alpha},[M^{\alpha},\,\cdot\,]\bigr]\Bigr)\,t^{0}
  \label{eq:DWop}
\end{equation}
which is the operator Debye--Waller factor of
Refs.~\cite{Fritzsche1994,Brouder,PoiarkovaRehr}.

Two remarks on the conventions. Isotropy is a specialisation and not a
requirement of the formalism: for general Gaussian vibrations the same
steps give
\begin{equation}
  \bar T=\exp\Bigl(-\tfrac12k^{2}\sum_{\alpha\beta}C_{\alpha\beta}
  \bigl[M^{\alpha},[M^{\beta},\,\cdot\,]\bigr]\Bigr)\,t^{0},
  \qquad
  C_{\alpha\beta}=\langle u_\alpha u_\beta\rangle,
  \label{eq:DWaniso}
\end{equation}
of which \eqref{eq:DWop} is the case
$C_{\alpha\beta}=\sigma^{2}\delta_{\alpha\beta}$; we adopt the isotropic
form only to keep the formulae readable. Second, the meaning of
$\sigma^{2}$ must be kept firmly in view, because the same symbol carries
a different meaning in the analysis of EXAFS data. Here $\sigma^{2}$ is
the variance of \emph{one Cartesian component} of the displacement of
\emph{one site}, so that $\langle|\vec u|^{2}\rangle=3\sigma^{2}$; there
it denotes the mean-square relative displacement of a \emph{pair} of
atoms projected on the bond direction,
$\langle(\vec u_{ij}\cdot\hat R_{ij})^{2}\rangle$. The two differ in being
a single-site rather than a relative quantity, and in the projection. All
the formulae of this section, including $d=k^{2}\sigma^{2}$ in
Sec.~\ref{sec:brouder}, are written in the first convention.

It is worth being
explicit about the step that produces \eqref{eq:DWop}, because the
notebook derivation compresses it. At order $2n$ the average of
$u_{\alpha_1}\cdots u_{\alpha_{2n}}$ is the sum over all Wick pairings,
each pairing contributing $\sigma^{2n}$ times a product of Kronecker
deltas; there are $(2n-1)!!$ of them, and they do \emph{not} all carry the
same index structure. They collapse onto the single expression
\eqref{eq:DWop} only because the generators commute, \eqref{eq:Mcomm}, so
that the nested commutators may be reordered at will; the combinatorial
factor $(2n-1)!!/(2n)!=1/(2^{n}n!)$ is exactly the coefficient of the
exponential.

Expanding \eqref{eq:DWop} order by order gives the recursion
\begin{equation}
  \bar T=\sum_{n}\bar T^{(n)},
  \qquad
  \bar T^{(n+1)}=-\frac{k^{2}\sigma^{2}}{2(n+1)}\sum_{\alpha}
  \Bigl(M^{\alpha}M^{\alpha}\bar T^{(n)}
  +\bar T^{(n)}M^{\alpha}M^{\alpha}
  -2M^{\alpha}\bar T^{(n)}M^{\alpha}\Bigr),
  \label{eq:Trec}
\end{equation}
with $\bar T^{(0)}=t^{0}$. Using the closure relation
\begin{equation}
  \sum_{\alpha}M^{\alpha}M^{\alpha}=\id
  \label{eq:closure}
\end{equation}
the first order reduces to
\begin{equation}
  \bar T^{(1)}=-k^{2}\sigma^{2}
  \Bigl(t^{0}-\sum_{\alpha}M^{\alpha}t^{0}M^{\alpha}\Bigr),
\end{equation}
and the remaining sum is evaluated with the identity
$\sum_\alpha X^{\alpha}Y^{\alpha}=X^{z}Y^{z}-X^{-}Y^{+}-X^{+}Y^{-}$ and
the Gaunt integrals, giving
\begin{equation}
  \bar T^{(1)}_{L'L}=-k^{2}\sigma^{2}
  \Bigl[t^{0}_{\ell}-\sum_{\ell''}(2\ell''+1)\,t^{0}_{\ell''}
  \begin{pmatrix}\ell&\ell''&1\\0&0&0\end{pmatrix}^{2}\Bigr]
  \delta_{L'L} .
  \label{eq:T1}
\end{equation}
Since $\bar T^{(1)}$ is diagonal in $L=(\ell,m)$, so are all the
$\bar T^{(n)}$. Introducing
\begin{equation}
  a_{\ell'\ell}=\ii^{\ell}\sqrt{2\ell+1}\,\delta_{\ell'\ell},
  \qquad
  m_{\ell'\ell}=\ii^{\,\ell'-\ell}\sqrt{(2\ell'+1)(2\ell+1)}
  \begin{pmatrix}\ell'&\ell&1\\0&0&0\end{pmatrix}^{2},
\end{equation}
one verifies $a^{-1}m\,a=(2\ell+1)\bigl(\begin{smallmatrix}
\ell'&\ell&1\\0&0&0\end{smallmatrix}\bigr)^{2}$ and the first order takes
the form $\bar T^{(1)}_{\ell'}=-k^{2}\sigma^{2}\sum_{\ell}
\{\id-a^{-1}m\,a\}_{\ell'\ell}t^{0}_{\ell}$. Iterating,
\begin{equation}
  \bar T_{\ell'}=\sum_{\ell}
  \Bigl\{\exp\bigl[-k^{2}\sigma^{2}\bigl(\id-a^{-1}m\,a\bigr)\bigr]
  \Bigr\}_{\ell'\ell}\;t^{0}_{\ell},
  \label{eq:fritzsche51}
\end{equation}
which is Eq.~(51) of Ref.~\cite{Fritzsche1994}. Because the similarity
transformation passes through the exponential,
$\exp(k^{2}\sigma^{2}a^{-1}m\,a)=a^{-1}\exp(k^{2}\sigma^{2}m)\,a$, an
identity valid for any invertible $a$, each term of the series
telescoping, and because the diagonal displacement matrix obeys
$J_{\ell'\ell}(ku)=\exp(\ii ku\,m_{\ell'\ell})$, one arrives at
\begin{equation}
  \bar T_{\ell'}=e^{-k^{2}\sigma^{2}}\sum_{\ell}
  \Bigl\{a^{-1}J\bigl(-\ii k^{2}\sigma^{2}\bigr)a\Bigr\}_{\ell'\ell}
  t^{0}_{\ell},
  \label{eq:fritzsche52}
\end{equation}
Eq.~(52) of the same work.

\subsection{The small-disorder limit: Brouder's formula}
\label{sec:brouder}

Equation \eqref{eq:fritzsche52} can be expanded in spherical Bessel
functions of increasing order. Writing $d=k^{2}\sigma^{2}$ and using
$j_0(0)=1$, $j_n(0)=0$ for $n>0$ together with the recurrence
$j_{n+1}(x)=\frac{2n+1}{x}j_n(x)-j_{n-1}(x)$, one obtains
\begin{equation}
  \bar T_{\ell'}=e^{-d}\Bigl\{
  \sum_{\ell}(2\ell+1)
  \begin{pmatrix}\ell'&0&\ell\\0&0&0\end{pmatrix}^{2}j_0(-\ii d)\,t^{0}_{\ell}
  +\ii\sum_{\ell}3(2\ell+1)
  \begin{pmatrix}\ell'&1&\ell\\0&0&0\end{pmatrix}^{2}j_1(-\ii d)\,t^{0}_{\ell}
  -\cdots\Bigr\}.
\end{equation}
The $3j$ symbol with a zero entry is non-vanishing only for $\ell=\ell'$,
where $\bigl(\begin{smallmatrix}\ell&\ell&0\\0&0&0\end{smallmatrix}\bigr)
=(-1)^{\ell}/\sqrt{2\ell+1}$, while the one containing a unit entry
requires $\ell=\ell'\pm1$. Retaining the first two terms and using
$j_0(-\ii d)\to1$, $3\ii j_1(-\ii d)\to d$ for small $d$,
\begin{equation}
    \bar T_{\ell}\simeq e^{-d}\left\{t_{\ell}
  +d\;\frac{\ell\,t_{\ell-1}+(\ell+1)\,t_{\ell+1}}{2\ell+1}\right\}
  \label{eq:brouder}
\end{equation}
which is the formula of Brouder~\cite{Brouder}. Two features of
\eqref{eq:brouder} anticipate the general result of
Sec.~\ref{sec:unified}: the correction couples only neighbouring angular
momenta, and it is controlled by how much the $t$-matrix varies from one
channel to the next.

\subsection{Thermal disorder in the augmented space}
\label{sec:thermaug}

So far the average has been performed by hand on a Gaussian distribution.
We now obtain the same result as a matrix element of a non-random
operator, which is what allows chemical and thermal disorder to be treated
together.

By Sec.~\ref{sec:blocks}, a Gaussian displacement of variance $\sigma^{2}$
is represented by the Jacobi matrix $R$ of Eq.~\eqref{eq:mgauss}, with
$b_n=\sigma\sqrt n$ and $\mel{f_0}{R^{n}}{f_0}=\mu_n$. Each site carries
three displacement variables $u_x,u_y,u_z$, hence three independent
configuration factors: for a single site
$\Phi^{(i)}=\varphi^{(ix)}\otimes\varphi^{(iy)}\otimes\varphi^{(iz)}$,
with $R_\alpha=\id\otimes\cdots\otimes R\otimes\cdots\otimes\id$ acting as
the Jacobi matrix on its own factor and as the identity on the others, and
with the product reference state
$\ket{F^{(i)}}=\ket{f_0^{(ix)}}\otimes\ket{f_0^{(iy)}}\otimes
\ket{f_0^{(iz)}}$; for the whole cluster
$\Phi=\bigotimes_{i\alpha}\varphi^{(i\alpha)}$ and
$\ket{F}=\bigotimes_{i\alpha}\ket{f_0^{(i\alpha)}}$. The symbol
$\ket{f_0}$ below stands for this product. Promoting the
displacement in \eqref{eq:Jexp},
\begin{equation}
  \mathcal{J}(R)=\exp\Bigl(\ii k\sum_{\alpha}R_\alpha\otimes M^{\alpha}\Bigr),
\end{equation}
the averaged $t$-matrix becomes
\begin{equation}
  \begin{split}
  \bar T&=\mel{f_0}{\mathcal{J}(-R)\,t^{0}\,\mathcal{J}(R)}{f_0}\\
  &=t^{0}-\ii k\sum_{\alpha}\mel{f_0}{R_\alpha}{f_0}
   \bigl[M^{\alpha},t^{0}\bigr]
  -\frac{k^{2}}{2}\sum_{\alpha\beta}
  \mel{f_0}{R_\alpha R_\beta}{f_0}\bigl[M^{\alpha},[M^{\beta},t^{0}]\bigr]
  +\cdots
  \end{split}
\end{equation}
The moments of the reference state are those of the Gaussian; since
operators belonging to different Cartesian factors commute, and
$\mel{f_0}{R^{2n}}{f_0}$ reproduces the moments of a centred Gaussian on
each factor, they obey Wick's theorem,
\begin{equation}
  \begin{aligned}
  &\mel{f_0}{R_\alpha}{f_0}=0,
  &&\mel{f_0}{R_\alpha R_\beta}{f_0}=\sigma^{2}\delta_{\alpha\beta},\\
  &\mel{f_0}{R_\alpha R_\beta R_\rho}{f_0}=0,
  &&\mel{f_0}{R_\alpha R_\beta R_\rho R_\theta}{f_0}
    =\sigma^{4}\bigl(\delta_{\alpha\beta}\delta_{\rho\theta}
    +\delta_{\alpha\rho}\delta_{\beta\theta}
    +\delta_{\alpha\theta}\delta_{\beta\rho}\bigr),
  \end{aligned}
\end{equation}
and so on, the moment of order $2n$ being the sum of the $(2n-1)!!$
pairings. The value $3\sigma^{4}$ is recovered when all four indices
coincide, $\mel{f_0}{R_x^{4}}{f_0}=3\sigma^{4}$, whereas mixed moments
such as $\mel{f_0}{R_x^{2}R_y^{2}}{f_0}=\sigma^{4}$ are non-zero and are
precisely what a delta requiring all four indices to be equal would
discard. Retaining the full pairing structure is not optional: it is what
makes the fourth-order term equal to $\tfrac18k^{4}\sigma^{4}
\sum_{\alpha\beta}\mathrm{ad}_{\alpha}^{2}\,\mathrm{ad}_{\beta}^{2}$, with
$\mathrm{ad}_\alpha\equiv[M^{\alpha},\,\cdot\,]$, and hence to the fourth
order of the exponential below. With the pairings in place the series
resums to
\begin{equation}
  \bar T=\exp\Bigl(-\tfrac12k^{2}\sigma^{2}\sum_{\alpha}
  \bigl[M^{\alpha},[M^{\alpha},\,\cdot\,]\bigr]\Bigr)\,t^{0},
\end{equation}
identical to \eqref{eq:DWop}. The thermal average is thus a matrix element
of a non-random operator on a reference state, exactly as the chemical one
in Eq.~\eqref{eq:tauaug}.

\paragraph{The bridge.}
It remains to carry the displacement operators through the inversion. The
$t$-matrices are site-diagonal and the propagators are not; collecting the
displacement matrices of all sites into the block-diagonal
$\mathcal{J}(\vec u)=\mathrm{diag}\{J(\vec u_0),\dots,J(\vec u_{N-1})\}$,
Eq.~\eqref{eq:Gshift} reads $G\to\mathcal{J}(\vec u)G\mathcal{J}(-\vec u)$
and therefore
\begin{equation}
  \bigl[T_a^{-1}+\mathcal{J}(\vec u)G\mathcal{J}(-\vec u)\bigr]^{-1}
  =\mathcal{J}(\vec u)\bigl[\widetilde{T}_a^{-1}+G\bigr]^{-1}
  \mathcal{J}(-\vec u),
  \qquad
  \widetilde{T}_a=\mathcal{J}(-\vec u)\,T_a\,\mathcal{J}(\vec u),
\end{equation}
where $\mathcal{J}^{-1}(\vec u)=\mathcal{J}(-\vec u)$ and
$[ABC]^{-1}=C^{-1}B^{-1}A^{-1}$ have been used. Taking the $00$ block,
\begin{equation}
    \tau^{00}=J(\vec u_0)\,
  \Bigl[\bigl(\id+\widetilde{T}_aG\bigr)^{-1}\Bigr]^{00}\,
  J(-\vec u_0)\;t^{0}
  \label{eq:bridge}
\end{equation}
All the dependence on the displacements has moved from the propagators
onto the scatterers, where the augmented-space average of the previous
paragraph applies directly. The two outer factors $J(\pm\vec u_0)$ can be
removed altogether: translational invariance allows the instantaneous
position of the absorber to be taken as the origin, so that $\vec u_0=0$,
$J(\vec u_0)=\id$ and $\bar T^{0}\equiv t^{0}$, provided the remaining
variables are understood as displacements \emph{relative} to the absorber,
$\vec v_i=\vec u_i-\vec u_0$. This is a choice of reference frame and not
a conditioning of the ensemble. It has no counterpart in
Sec.~\ref{sec:crosssec}: there the occupation of the origin is fixed by
the physics of the edge, whereas here the absorber does vibrate, and what
is being used is only the invariance of the spectrum under a rigid
displacement of the whole cluster.

We can now write the thermal counterpart of \eqref{eq:tauaug}, which is
the actual object of the theory. Promoting each relative displacement to
its Jacobi operator, $\vec v_i\to\vec R_i$, the scatterers become
\begin{equation}
  \widetilde{\mathcal{T}}^{\,i}
  =\mathcal{J}(-\vec R_i)\,t^{i}\,\mathcal{J}(\vec R_i),
  \qquad
  \bigl(\widetilde{\mathcal{T}}^{\,i}\bigr)^{-1}
  =\mathcal{J}(-\vec R_i)\,\bigl(t^{i}\bigr)^{-1}\,\mathcal{J}(\vec R_i),
  \label{eq:Ttherm}
\end{equation}
the second identity following from the first because the promotion is a
similarity transformation, the thermal analogue of the spectral argument
that gave \eqref{eq:affine} for chemical disorder, and, as there, the
reason why promoting $t$ and promoting $t^{-1}$ are simultaneously exact.
The augmented secular matrix is
\begin{equation}
  \mathcal{A}_{\rm th}
  =\widetilde{\mathcal{T}}_a^{-1}\bigl(\{\vec R_i\}\bigr)+G,
  \label{eq:Ath}
\end{equation}
with $\widetilde{\mathcal{T}}_a$ block-diagonal and $G$ evaluated between
the \emph{equilibrium} positions, and the configuration average of the
scattering-path operator is once more a single matrix element of an
inverse,
\begin{equation}
    \bigl\langle\tau^{00}\bigr\rangle
  =\mel{0F}{\mathcal{A}_{\rm th}^{-1}}{0F},
  \qquad
  \ket{F}=\bigotimes_{i\alpha}\ket{f_0^{(i\alpha)}}
  \label{eq:tauaugth}
\end{equation}
This, and not \eqref{eq:DWop}, is the result of the section. The averaged
$t$-matrix is a by-product: substituting $\bar T$ for every scatterer
would reproduce $\langle\tau^{00}\rangle$ only if the average factorised
along the path, which is precisely what fails for the paths examined in
Sec.~\ref{sec:unified}.

\paragraph{Correlated displacements.}
As it stands the construction assigns three statistically independent
Gaussian variables to every site. In this site-factorised Cartesian
representation the construction is exact for independent local
oscillators of the Einstein type. In a harmonic solid the displacements of different sites are
correlated, $\langle u_{i\alpha}u_{j\beta}\rangle\neq0$, and referring the
displacements to the absorber introduces correlations even when the
absolute ones are independent, since $\langle v_{i\alpha}v_{j\beta}\rangle
=\langle(u_{i\alpha}-u_{0\alpha})(u_{j\beta}-u_{0\beta})\rangle$.

The formalism accommodates this without any modification, provided the
independent variables are chosen to be the normal coordinates rather than
the Cartesian displacements. Writing
$u_{i\alpha}=\sum_{\nu}e_{i\alpha,\nu}q_{\nu}$, with $e$ the polarisation
vectors and $q_\nu$ the normal coordinates, statistically independent and
Gaussian, with variances fixed by the mode frequencies and the
temperature~\cite{PoiarkovaRehr}, one promotes each $q_\nu$ to its own
Jacobi operator $R_\nu$, so that $\Phi=\bigotimes_\nu\varphi^{(\nu)}$ and
the displacement entering site $i$ becomes $\sum_\nu e_{i\alpha,\nu}
R_\nu$. Correlations between sites, and those generated by referring the
displacements to the absorber,
$v_{i\alpha}=\sum_\nu(e_{i\alpha,\nu}-e_{0\alpha,\nu})q_\nu$, are then
carried automatically by the polarisation vectors, and
\eqref{eq:tauaugth} holds unchanged.

The price is computational rather than formal, and should be stated. In
the Cartesian description $\mathcal{J}(\vec R_i)$ acts on the three
factors belonging to site $i$ alone, and $\mathcal{A}_{\rm th}$ inherits
the same one-site-per-factor sparsity that keeps the recursion of
Appendix~\ref{app:cf} cheap. In the normal-mode description every
$\mathcal{J}$ acts on all $3N$ mode factors at once and that sparsity is
lost. Which description is preferable is thus a question about the system
rather than about the formalism: independent local oscillators where the
correlations are weak, normal modes where they are not.

\section{Unified treatment and the repeated-site correction}
\label{sec:unified}

Everything up to this point has been exact. Equations~\eqref{eq:tauaug}
and \eqref{eq:tauaugth} deliver $\langle\tau^{00}\rangle$ as a matrix
element of an inverse, and for a calculation carried out in the augmented
space nothing further is required: the results of this section are already
contained in them. The question asked here is a different one, and its
answer is useful precisely to those who do \emph{not} build the augmented
space. When explicit configurational sampling is not used, the standard
analytic prescription, in both extended- and near-edge analyses, is to
retain an ordinary multiple-scattering calculation and to replace each
scatterer by its configuration average. We ask when
that prescription is exact, and what the leading error is when it is not.
The answer is a closed-form correction which we obtain from the augmented
space but which is expressed entirely in terms of quantities available in
an ordinary calculation, so that it can be applied without constructing
the augmented space.

The first half of the answer is already in hand. Whenever the disorder
variables attached to distinct sites are statistically independent and
every site of a
scattering path occurs once, the average of the product is the product of
the averages, and the whole effect of the disorder is carried by the
averaged $t$-matrices: $\bar t=y\,t^{B}+x\,t^{A}$ for chemical disorder,
$\bar T$ of Eq.~\eqref{eq:DWop} for thermal disorder, and the composition
of the two when both are present. This is what Eqs.~\eqref{eq:chi2} and
\eqref{eq:chi3} showed explicitly for single and third-order scattering,
and the same argument holds at every order; it is also the assumption
underlying the standard analysis of experimental data. The assumption
fails as soon as a site appears twice in the same path. It fails also, and
for an independent reason, when the variables of distinct sites are
correlated, as they are once the displacements are referred to the
absorber and, in general, in the normal-mode description of
Sec.~\ref{sec:thermaug}: two scatterers visited once each then depend on
common variables, and the average of their product keeps a covariance
term. Equation~\eqref{eq:tauaugth} contains that case without
approximation, since the variables promoted there are the independent
ones. What fails is only the path-by-path factorisation into averaged
scatterers, and the closed correction derived below addresses the first
mechanism.

\subsection{The repeated-site correction}
\label{sec:Dcorr}

Consider a scattering path that comes back to a site $i$ it has already
visited. The two visits carry the \emph{same} disorder variable, so that
the average of the product of the two $t$-matrices is not the product of
the averages, and something survives. To identify what survives we do not
follow one path: we perturb the scatterer at site $i$ about its
configuration average and expand $\tau=(T_a^{-1}+G)^{-1}$ to second order.
Since only the site-$i$ block of $T_a^{-1}$ depends on $t^{i}$, the
expansion closes on three objects,
\begin{equation}
  \Pi^{ii}=\bar t^{\,-1}\bigl(\tau^{ii}-\bar t\,\bigr)\bar t^{\,-1},
  \qquad
  L^{0i}=\tau^{0i}\,\bar t^{\,-1},
  \qquad
  L^{i0}=\bar t^{\,-1}\tau^{i0},
  \label{eq:Pidef}
\end{equation}
all evaluated on the cluster in which every scatterer carries its average.
$\Pi^{ii}$ is the whole propagation that leaves site $i$ and returns to
it, with the two scattering events at its ends amputated; they are
supplied by the correction itself. $L^{0i}$ and $L^{i0}$ are the
correspondingly amputated legs joining the absorber to that site. The
configuration average of the scattering-path operator is then
\begin{equation}
  \bigl\langle\tau^{00}\bigr\rangle
  =\tau^{00}\bigl[\bar t\,\bigr]
  +\sum_{i}L^{0i}\,D^{i}\,L^{i0}+\dots,
  \label{eq:tauexp}
\end{equation}
the sum running over the disordered sites. The order of the terms hidden
in the dots depends on which disorder one is treating. For Gaussian
displacements the odd moments vanish and the next contribution is of
fourth order. For a binary occupation this holds only at equiatomic
composition: the third central moment of $n_i$ is
\begin{equation}
  \bigl\langle(n_i-x)^{3}\bigr\rangle=xy\,(y-x),
  \label{eq:thirdmom}
\end{equation}
which vanishes if and only if $x=y=\tfrac12$. Away from that composition a
third-order term is present, and it is the first one neglected.

It is worth saying plainly what \eqref{eq:tauexp} is and what it is not.
It is the exact coefficient of the second order in the disorder, not the
complete correction at arbitrary contrast: the expansion parameter is
$\Delta t$ in the chemical case and $k\sigma$ in the thermal one, and the
truncation is controlled only while those are small. The numerical test of
Sec.~\ref{sec:validation} is built accordingly: the ratio of the measured
correction to the predicted one tends to unity as the contrast between the
two scatterers is reduced. It therefore establishes the coefficient,
not the adequacy of a second-order treatment for a strongly contrasted
alloy, where the higher orders would have to be retained.

For thermal disorder the site correction is
\begin{equation}
  D^{i}=-k^{2}\sigma^{2}\sum_{\alpha}
    \bigl[M^{\alpha},t^{i}\bigr]\,\Pi^{ii}\,
    \bigl[M^{\alpha},t^{i}\bigr]
   +O\bigl(k^{4}\sigma^{4}\bigr),
  \label{eq:Ddiag}
\end{equation}
obtained by expanding to second order in the displacement: the term linear
in the displacement averages to zero, and what survives at second order is
the average of the product of two first-order terms, which does not
factorise.

It matters that $\Pi^{ii}$ and the legs in \eqref{eq:Pidef} are the
\emph{full} quantities and not their lowest-order approximations
$\tau^{ii}-\bar t\simeq\bar t\,G^{iJ}\bar t\,G^{Ji}\bar t$ and
$\tau^{0i}\simeq-t^{0}G^{0i}\bar t$, which would retain only the shortest
path that revisits the site, $0\to i\to J\to i\to0$. Every other path that
passes twice through site $i$, reaching it indirectly, or wandering
further between the two visits, carries the same second moment of the
same disorder variable and therefore contributes at the same order. On a
compact cluster these are not a small correction to the shortest path but
comparable to it, as the test of Sec.~\ref{sec:validation} shows.

The sum over the three Cartesian components in \eqref{eq:Ddiag} is a
\emph{diagonal} sum: it descends from
$\langle u_\alpha u_\beta\rangle=\sigma^{2}\delta_{\alpha\beta}$ for
isotropic vibrations. It does not factorise into a single commutator with
$A=\sum_\alpha M^{\alpha}$, because the terms with $\alpha\neq\beta$ are
precisely those the Kronecker delta removes. The three components must
therefore be kept separate. Using
\begin{equation}
  \sum_{\alpha=x,y,z}X^{\alpha}Y^{\alpha}
  =X^{z}Y^{z}-X^{-}Y^{+}-X^{+}Y^{-},
  \label{eq:cart2sph}
\end{equation}
which is the same pattern of signs that produces the closure relation
\eqref{eq:closure}, one has
\begin{equation}
  D=-k^{2}\sigma^{2}\Bigl\{
    [M^{z},t]\,\Pi\,[M^{z},t]
   -[M^{-},t]\,\Pi\,[M^{+},t]
   -[M^{+},t]\,\Pi\,[M^{-},t]\Bigr\}.
  \label{eq:Dsph}
\end{equation}

Each $M^{\mu}$ is a tensor operator of rank one, so it obeys on its own
the selection rules $\ell=\ell'\pm1$ and $m=m'+\mu$: the three components
transfer $m$ by $-1$, $0$ and $+1$ and live in disjoint blocks, which is
the structural reason why they cannot be collected into a single operator.
With $t$ diagonal in $L$ the commutator is elementary,
\begin{equation}
  \bigl[M^{\mu},t\bigr]_{LL'}=M^{\mu}_{LL'}\,\bigl(t_{\ell'}-t_{\ell}\bigr),
\end{equation}
and, introducing the differences of consecutive $t$-matrix elements
$\Delta_{\ell}\equiv t_{\ell}-t_{\ell+1}$,
\begin{equation}
    \begin{aligned}
  \bigl[M^{\mu},t\bigr]_{LL'}
  &=\ii\Bigl[\,b^{+}_{\ell'}\,\Delta_{\ell'}\,\delta_{\ell,\ell'+1}
   -b^{-}_{\ell'}\,\Delta_{\ell'-1}\,\delta_{\ell,\ell'-1}\Bigr]
   \,C^{\ell m}_{\ell'm',1\mu},\\[4pt]
  b^{+}_{\ell'}&=\sqrt{\frac{\ell'+1}{2\ell'+3}},\qquad
  b^{-}_{\ell'}=\sqrt{\frac{\ell'}{2\ell'-1}}
  \end{aligned}
  \label{eq:comm-explicit}
\end{equation}
where $C^{\ell m}_{\ell'm',1\mu}$ are Clebsch--Gordan
coefficients~\cite{Varshalovich}. Two consequences are now manifest.
First, the correction depends on the $t$-matrix only through the
differences $\Delta_{\ell}$ between consecutive angular momenta; second,
and as a corollary, $D$ vanishes identically for a scatterer whose
$t$-matrix does not depend on $\ell$. The size of $\Delta_\ell$ in a
simple case is obtained in Appendix~\ref{app:born}.

\subsection{Chemical disorder and the unified form}
\label{sec:unified-form}

The same mechanism operates for substitutional disorder, and is even
simpler to exhibit. A path that passes twice through site $i$ requires
$\langle n_i^{2}\rangle$, whereas the factorised prescription supplies
$\langle n_i\rangle^{2}$; since $n_i$ is a binary variable,
$\langle n_i^{2}\rangle=\langle n_i\rangle=x$ and the mismatch is
$x-x^{2}=xy$. In operator language it is the projector property
\eqref{eq:proj} that produces it: the two factors $M^{(i)}$ collapse to
one instead of factorising into $\mel{F}{M^{(i)}}{F}^{2}$. The site correction to be
inserted in \eqref{eq:tauexp} is therefore
\begin{equation}
  D^{i}_{\text{chem}}=xy\;\Delta t\;\Pi^{ii}\;\Delta t,
  \qquad \Delta t=t^{A}-t^{B},
  \label{eq:Dchem}
\end{equation}
which is quadratic in the difference between the two scatterers and
vanishes, as it must, for $x\to0$ or $x\to1$.

Equations \eqref{eq:Ddiag} and \eqref{eq:Dchem} are the same statement.
Both express the failure of the average of a square to equal the square of
the average, and both are governed by the variance of the underlying
disorder variable times the sensitivity of the $t$-matrix to it,
\begin{equation}
    D^{i}=\sum_{\xi}\mathrm{Var}(\xi)\;
  \frac{\partial t}{\partial\xi}\;\Pi^{ii}\;
  \frac{\partial t}{\partial\xi}
  \label{eq:Dunified}
\end{equation}
the sum running over the independent disorder variables of the site: the
occupation number, with $\mathrm{Var}(n)=xy$ and
$\partial t/\partial n=\Delta t$, and the three displacement components,
with $\mathrm{Var}(u_\alpha)=\sigma^{2}$ and
$\partial t/\partial u_\alpha=-\ii k[M^{\alpha},t]$, as
\eqref{eq:tu} gives. The overall sign in
\eqref{eq:Ddiag} follows from the factor $\ii^{2}$.

This is the point at which the augmented-space treatment departs from the
standard analysis. A path in which every site occurs once is described
exactly by averaged $t$-matrices when the site variables are independent;
a path that revisits a site is not, and
the defect is a term that no rescaling of the $t$-matrices can reproduce,
since it involves the second moment of the disorder rather than the first.
Such paths are the ones that fold back on themselves, and they acquire
weight in the outer coordination shells, where multiple scattering is rich
and where, in oxides, substituted compounds and solutions no less than in
alloys, substitutional disorder is typically located.

\section{Anharmonic disorder}
\label{sec:anharm}

Nothing in the construction of Sec.~\ref{sec:thermaug} required the
displacement distribution to be Gaussian. What was used is that the
Gaussian is represented, through the inverse problem of
Sec.~\ref{sec:inverse}, by the Jacobi matrix \eqref{eq:mgauss} with
$a_n=0$ and $b_n=\sigma\sqrt n$. A non-Gaussian $p(u)$ is handled by the
same machinery: one computes its moments, forms the Hankel determinants
\eqref{eq:hankeldef} and reads off the coefficients \eqref{eq:hankel}. The
difference is that the diagonal coefficients no longer vanish and the
off-diagonal ones are no longer $\sigma\sqrt n$.

The leading effect of an asymmetric distribution is controlled by the
third cumulant. For a distribution
with $\mu_1=0$, $\mu_2=\sigma^{2}$ and $\mu_3=C_3$, the first two Hankel
determinants give
\begin{equation}
  a_0=0,\qquad a_1=\frac{C_3}{\sigma^{2}},
  \qquad\text{that is}\qquad a_1-a_0=\frac{C_3}{\sigma^{2}},
  \label{eq:anharm}
\end{equation}
so that the asymmetry of the single-site displacement distribution enters the disorder
operator as a shift of the diagonal of the Jacobi matrix, proportional to
the third cumulant, or equivalently to $\sigma$ times the dimensionless
skewness. Higher cumulants populate $a_2,a_3,\dots$ in the same way.
The off-diagonal sequence is affected as well, and only the first of its
entries is protected: $b_1^{2}=D_1=\mu_2=\sigma^{2}$ holds for any
distribution, whereas already
\begin{equation}
  b_2^{2}=\frac{D_2D_0}{D_1^{2}}
  =\frac{\mu_2\mu_4-\mu_3^{2}-\mu_2^{3}}{\mu_2^{2}}
  \label{eq:b2gen}
\end{equation}
reduces to $2\sigma^{2}$ only when $\mu_3=0$ and $\mu_4=3\sigma^{4}$, that
is only for the Gaussian. Both sequences of Jacobi coefficients, not the
diagonal one alone, carry the departure from Gaussian statistics.

Which of the two sequences carries the effect is decided by the symmetry
of the site. The distribution entering $\mathcal{J}(R)$ is that of the
displacement of a single atom, and when the site lies at a centre of
inversion the local potential is even: all odd moments vanish, every $a_n$
is zero, and the anharmonicity resides entirely in the $b_n$. This is what
one finds explicitly in the effective-potential calculations of Van Hung
and Fornasini~\cite{VanHungFornasini}, where the cubic term of the
single-atom effective potential of an fcc metal vanishes once the
interactions with all nearest neighbours are summed, while the
corresponding \emph{pair} potential remains strongly anharmonic. The two
are not interchangeable: the EXAFS cumulants of that work describe the
distribution of the absorber--backscatterer distance, in the pair
convention discussed in Sec.~\ref{sec:DW}, whereas the disorder operator
of a site is built from the moments of that site's own displacement.

The practical consequence is that anharmonicity requires no separate
theory and no cumulant expansion of the spectrum: the operator
$\mathcal{J}(R)$ is built from a Jacobi matrix with $a_n\neq0$, the
average is still the matrix element \eqref{eq:tauaugth}, and it remains
exact. What is lost is only the closed form \eqref{eq:DWop}, which relied
on the Gaussian resummation; the recursion of Appendix~\ref{app:cf} is
unaffected, since it assumes neither the Gaussian values of the $a_n$
nor those of the $b_n$.

\section{Numerical validation}
\label{sec:validation}

The formalism rests on two statements: the identity \eqref{eq:tauaug},
which turns the configuration average into a single matrix element, and
the second-order correction \eqref{eq:Dunified}, which gives the leading
contribution from paths that revisit a site. Both can be checked directly on clusters small enough for the
average to be performed by brute force. The scatterers are square wells of
radius $R_{MT}=1.20$~\AA{}, with phase shifts obtained by matching at the
well radius, no self-consistency, no relativistic corrections, nothing
that could obscure what is being tested. The propagators are the free ones
in the normalisation of Sec.~\ref{sec:crosssec}.

\paragraph{The augmented-space identity.}
We take the cluster written out in Sec.~\ref{sec:promotion}: the absorber
and two disordered neighbours, placed at the vertices of an equilateral
triangle of side $2.885$~\AA{} so that the two neighbours are nearest
neighbours of one another as well. The four occupations are summed with
their binomial weights and the result compared with the single matrix
element $\mel{0F}{\mathcal{A}^{-1}}{0F}$ of the $12\times12$ operator
\eqref{eq:A12}. The two agree to between $6\times10^{-17}$ and
$2\times10^{-16}$ at eight energies between $2$ and $5.5$~\AA$^{-1}$, that
is to machine precision.

It should be stressed what this does and does not verify. The identity
\eqref{eq:tauaug} holds for \emph{any} operator of the form
\eqref{eq:Aaug}, whatever the physical content of its blocks; the test
validates the augmented-space construction, not the choice of potentials
or of scattering conventions. That is precisely why a caricature of a
scatterer is adequate here, and why nothing would be gained by replacing
it with a realistic one.

\paragraph{The repeated-site correction.}
On the same cluster we compare the exact average with the prescription
that replaces every $t$-matrix by $\bar t$, and confront the difference
with the correction \eqref{eq:Dchem}. As the contrast $\Delta t$ is
reduced the ratio of the measured difference to the predicted one
approaches unity, $1.0042$, $0.9973$, $0.9998$, $1.0000$ for well-depth
differences of $0.30$, $0.08$, $0.02$ and $0.01$~Ha, confirming that
\eqref{eq:Dchem} is the exact coefficient of second order in the disorder.
Had $\Pi^{ii}$ been taken in its shortest-path form, the same ratio would
have converged to $1.83$ instead: on this compact cluster the paths that
reach the repeated site indirectly, or wander further between the two
visits, contribute as much as the shortest one.

\begin{figure}[htb]
  \centering
  \includegraphics[width=0.86\textwidth]{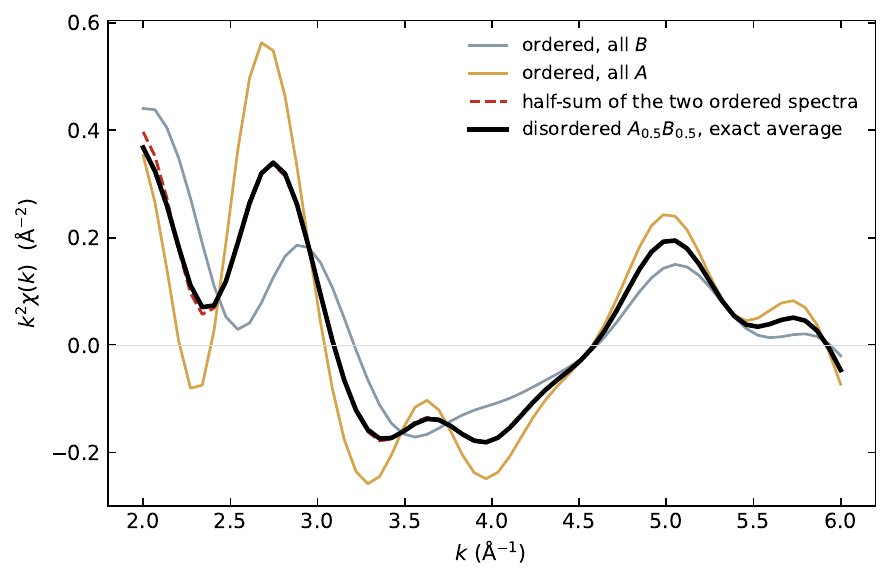}
  \caption{K-edge $k^{2}\chi(k)$ of a model tetrahedral cluster: an
  absorber surrounded by four neighbours at $2.50$~\AA{}. Thin lines: the
  two ordered clusters, all neighbours $B$ and all neighbours $A$. Thick
  line: the substitutionally disordered cluster $A_{0.5}B_{0.5}$, obtained
  as the exact average over the sixteen occupations and reproduced by the
  augmented-space identity \eqref{eq:tauaug} with a single inversion.
  Dashed line: the half-sum of the two ordered spectra, which is not the
  configuration average and departs from it by up to $8\%$ of the
  amplitude. The scatterers are square wells of radius $1.20$~\AA{}; the
  resulting phase shifts at $k=3$~\AA$^{-1}$ are
  $\delta_{0},\dots,\delta_{3}=0.53,\,0.60,\,0.50,\,0.14$ for $B$ and
  $0.90,\,0.88,\,0.90,\,0.30$ for $A$. The model carries no inelastic
  losses, which is why the structure below $k\simeq2.5$~\AA$^{-1}$ is
  sharper than in a measured spectrum.}
  \label{fig:disorder}
\end{figure}

\paragraph{A disordered cluster.}
Figure~\ref{fig:disorder} shows what the construction delivers on a
slightly larger system: an absorber surrounded by four neighbours at the
vertices of a tetrahedron, each of them $A$ or $B$ with equal probability.
The two ordered clusters bracket the disordered one, as they should, but
the spectrum of the alloy is not their interpolation: the half-sum of the
two ordered spectra departs from the true configuration average by up to
$8\%$ of its amplitude. The reason is that the average runs over
configurations, not over spectra, and the mixed configurations contain
scattering paths, an $A$ neighbour followed by a $B$ neighbour within the
same path, that neither ordered spectrum contains. It is exactly these
that the augmented space accounts for, and it does so with one inversion
in place of sixteen.

The magnitude of that $8\%$ is not a prediction: it is fixed by the
contrast between the two scatterers, which is a parameter of the model.
What does not depend on it is that the departure exists and is not
reducible to an interpolation between the end members.

\paragraph{Recursion depth.}
The identity \eqref{eq:tauaug} says nothing about how deep the continued
fraction must be carried, and that is what decides the calculation in
practice. We have measured it on a family of clusters consisting of an
absorber and $N$ disordered neighbours placed on a shell of radius
$2.5$~\AA{} at spherical Fibonacci points, so that the geometry is fixed
uniquely by $N$, with the same square-well scatterers as above,
$\ell_{\max}=2$, and $x=y=\tfrac12$. For each cluster the recursion was
run on $\mathcal{A}$ from the starting vector $\ket{0F}$ and the truncated
$J$-fraction monitored until it ceased to change by more than one part in
$10^{10}$. Between six and thirteen levels are required, the number
depending on the photoelectron energy, thirteen at $k=2$~\AA$^{-1}$,
seven at $k=4.5$~\AA$^{-1}$, but not on the size of the configuration
space: over ranks $(N+1)(\ell_{\max}+1)^{2}2^{N}$ ranging from
$7\times10^{2}$ to $10^{7}$, that is from $N=4$ to $N=16$, the number of
levels is flat to within one or two.

The behaviour is the expected one. The levels of the recursion are the
orders of the multiple-scattering series, Eq.~\eqref{eq:b1}, so their
number is set by how far the electron effectively propagates, which is a
property of the geometry and the energy; adding a disordered site adds one
scatterer to the cluster while doubling the configuration space, and the
two have no reason to grow together. We note also that full
reorthogonalisation of the Lanczos vectors, in the bilinear sense
appropriate to a complex symmetric operator, left every result unchanged
to machine precision: at these depths the recursion shows no sign of the
degradation that would call for a look-ahead step.

\section{Conclusions}
\label{sec:concl}

We have shown that the configuration average of the X-ray absorption cross
section of a disordered system can be performed exactly, once and for all,
at the level of the operators. The multiple-scattering expression of the
cross section is linear in the scattering-path operator, so the average of
the spectrum is the spectrum of the averaged scattering-path operator;
promoting the disorder variables to the operators of Mookerjee's augmented
space turns that average into an ordinary matrix element of a non-random
resolvent, which a continued-fraction recursion evaluates without ever
constructing the exponentially large space explicitly. The average is
obtained directly from that resolvent, without truncating a
scattering-path series: the construction delivers
$\langle\tau^{00}\rangle$ within the same full multiple-scattering scheme
in which near-edge spectra are computed, and its evaluation does not
require a scattering-path expansion.

Four points seem to us worth emphasising. First, the average is obtained
by inversion rather than through a scattering-path expansion. The disorder is built into the secular
matrix, which is then inverted exactly, so that nothing in the
construction presupposes the convergence of the multiple-scattering
series. This is what makes it usable in the near-edge region, where that
series does not converge and where the analytic treatments of disorder
developed for EXAFS, which introduce the damping path by path, cannot be
carried over. Explicit sampling also avoids the expansion, but it performs
the average numerically, over spectra, one full calculation per
configuration; here the average is analytic, is performed once, on the
scattering operator itself, and is exact.

Second, chemical and thermal
disorder are not two problems but one: they differ only in the dimension
of the local disorder operator, two for a binary occupation, infinite for a
Gaussian displacement, and the same promotion, the same reference state
and the same recursion serve both. The operator Debye--Waller factor of
the harmonic theory is recovered as the Gaussian special case, and
anharmonicity requires only that the diagonal and off-diagonal
coefficients of the Jacobi matrix be constructed from the non-Gaussian
moments.

Third, the analysis of scattering paths is a diagnostic applied to the
construction, not the route by which the average is obtained, and it
settles a question that the standard treatment leaves open. For
statistically independent site variables, the customary
prescription of replacing each scatterer by its configuration average is
exact for scattering paths that visit every site once, and only for
those. For a path that revisits a site, the leading
departure from the averaged-scatterer prescription is the exact
second-order term, Eq.~\eqref{eq:Dunified}. It is a bilinear form in the
derivative of the $t$-matrix, weighted by the variance of the disorder
variable and by the amputated return propagator. For substitutional
disorder it is
$xy\,\Delta t\,\Pi\,\Delta t$; for thermal disorder it is governed by the
differences $t_\ell-t_{\ell+1}$ between consecutive angular-momentum
channels and vanishes identically for a scatterer that does not resolve
them.

Fourth, the natural domain of the correction is not the first coordination
shell. Paths that revisit a site are the folded ones, and they gain weight
where multiple scattering is rich, that is in the outer shells, which is
also where substitutional disorder most commonly resides, in oxides, doped
and substituted compounds, and coordination environments in solution, no
less than in metallic alloys.

A word on cost, stated with the care the question deserves, since the
augmented space is exponentially large. Two facts are relevant, and they
bear on different halves of the calculation.

The first concerns the depth of the recursion. In test calculations on
clusters of up to sixteen disordered neighbours, the continued fraction
stabilises to ten significant figures after between six and thirteen
levels, the number depending on the photoelectron energy but not on the
size of the configuration space: over a range of ranks from
$7\times10^{2}$ to $10^{7}$ it is essentially flat. This is what one should expect, because
the levels of the recursion are the orders of the multiple-scattering
series, Eq.~\eqref{eq:b1}, and adding a disordered site adds one scatterer
to the geometry while doubling the configuration space. The depth is set
by the scattering order, not by the dimension of the space.

The second concerns the cost per level. Here the exponential factor is
real: a vector of the augmented space has $\mu^{N}$ components. It is not,
however, explored. Each term of $\mathcal{A}$ acts on a single
configuration factor, and the propagator does not act on $\Phi$ at all, so
that a Krylov vector at level $m$ has strictly vanishing amplitude on
every configuration differing from $\ket{F}$ in more than
$\lfloor m/2\rfloor$ sites, two applications of $\mathcal{A}$ being needed
to excite each new factor, one to reach the site and one to act on it,
the absorber itself carrying no disorder operator. The recursion is
therefore exactly confined, with no approximation, to a subspace whose
dimension, for binary substitutional disorder, is
$(N+1)(\ell_{\max}+1)^{2}\sum_{j\le\lfloor m/2\rfloor}\binom{N}{j}$ and
which for fixed $m$ grows polynomially in $N$; the corresponding bounds
for higher local rank and for continuous disorder are given in
Appendix~\ref{app:cf}. Taken
together with the flatness of $m$, this is the reason to expect the method
to remain viable where enumeration cannot.

We are deliberately not claiming more than that. The clusters on which we
have tested the construction are small enough that the confinement buys
little (at sixteen neighbours it is a factor of a few), and the regime in
which it would matter is one we have not computed. An implementation that
exploits it, and a comparison against enumeration on a real material, are
a separate undertaking, and the present work is intended as the
specification for one rather than a substitute for it.

A development suggests itself. Throughout this work the site occupations
were taken to be statistically independent, and the reference state
$\ket{F}$ was accordingly a simple product. Short-range order breaks that
independence. Short-range order can, in principle, be encoded in a
correlated reference state while retaining the same local occupation
operators: it does not modify the operator $\mathcal{A}$ at all, moving
entirely into $\ket{F}$, which then ceases to factorise and carries the
Warren--Cowley parameters as its correlations. Fixing the absorber, which
we imposed in Sec.~\ref{sec:crosssec} on physical grounds, is itself the
simplest instance of such a conditioning. Since the two edges of a binary
system are two distinct calculations rather than two components of one
average, the comparison of the two absorption edges can then carry
information on the local chemical
order. This is a natural direction for future work.

What we believe has been settled here is a question of principle. The
configuration average of a core-level absorption spectrum, over chemical
substitution and thermal motion together, can be carried out exactly, at
the level of the operators and before any spectrum is computed, without
leaving the full multiple-scattering scheme, and the object it produces is
a single matrix element of a non-random resolvent.
No effective medium is introduced, no self-consistency is required, no
form is assumed for the distributions, and the averaged-scatterer
prescription is recovered as the leading term of a controlled expansion
rather than postulated. The construction applies to one disordered species
or several, and to harmonic or anharmonic vibrations; it can in principle
be extended to correlated occupations by allowing the reference state to
carry the corresponding correlations.

What has not been settled is everything that follows from it in practice.
We have verified the identity and the correction to machine precision on
clusters small enough to be enumerated explicitly, and we have
established how
the recursion behaves as the space grows; we have not built an
implementation, nor confronted the method with a measured spectrum. Those
steps will decide whether the construction becomes a working tool, and
they are the natural continuation of this work. We have tried to write it
so that they can be taken by others as readily as by us.

\section*{Acknowledgements}

It is a pleasure to thank C.\ R.\ Natoli (INFN, Laboratori Nazionali di
Frascati), with whom I have worked on the theory of X-ray absorption for
the better part of a working life, and from whom most of what I understand
of multiple scattering was learnt. I am equally indebted to P.\ Frank
(Stanford Synchrotron Radiation Lightsource) for his encouragement while
this work was being written, and for persuading me that it was worth
publishing at all. To both I owe a great deal.

\medskip
\noindent
The author declares no conflict of interest. This work received no
specific grant from any funding agency in the public, commercial or
not-for-profit sectors.

\section*{Data availability statement}

No new data were created or analysed in this study. The work is
theoretical, and all the information needed to reproduce the numerical
checks reported in Sec.~\ref{sec:validation} and in
Fig.~\ref{fig:disorder}, namely the cluster geometries, the square-well
parameters, the angular-momentum cut-off and the energy range, is
contained within the article.

\section*{Declaration of generative AI use}

In preparing this work the author made use of a large language model
(Claude, Anthropic) as an assistant in editing the
manuscript and in carrying out numerical
verifications. The theoretical content is the author's own work and is based on a derivation elaborated in unpublished notes written in 2013. The author has reviewed and edited the entire text and assumes full responsibility for the content of the publication.

\appendix

\section{Proof of the augmented-space theorem}
\label{app:proof}

Let $M$ be an operator on $\phi$ with spectral decomposition
$M=\sum_J d_J P_J$, $d_J$ its eigenvalues, and let
\begin{equation}
  E_\lambda=\sum_{d_J\le\lambda}P_J
\end{equation}
be its spectral family. Then $M$ admits the Stieltjes representation
\begin{equation}
  M=\int\lambda\,\dd E_\lambda,\qquad
  \dd E_\lambda=E_\lambda-E_{\lambda-0},
\end{equation}
whence, for any function $f$,
\begin{equation}
  f(M)=\int f(\lambda)\,\dd E_\lambda,\qquad
  \mel{v}{f(M)}{v'}=\int f(\lambda)\,
  \dd\!\braket{v}{E_\lambda\,v'},
  \label{eq:stieltjes}
\end{equation}
the last quantity being an ordinary scalar measure. In particular the
resolvent has the representation
\begin{equation}
  \Res(z)=\int\frac{\dd E_\lambda}{z-\lambda} .
\end{equation}
Using the definition \eqref{eq:pdos} together with
\begin{equation}
  \frac{1}{x-x_0+\ii\varepsilon}
  =\mathcal{P}\left(\frac{1}{x-x_0}\right)-\ii\pi\,\delta(x-x_0),
\end{equation}
one has
\begin{equation}
  p(x)=-\frac1\pi\lim_{\varepsilon\to0^{+}}\mathrm{Im}\int
  \frac{\dd\!\braket{v_0}{E_\lambda\,v_0}}{x+\ii\varepsilon-\lambda}
  =\int\delta(x-\lambda)\,\dd\!\braket{v_0}{E_\lambda\,v_0},
\end{equation}
so that for any $f$
\begin{equation}
  \int f(x)\,p(x)\,\dd x
  =\int f(\lambda)\,\dd\!\braket{v_0}{E_\lambda v_0}
  =\mel{v_0}{f(M)}{v_0} .
  \label{eq:key}
\end{equation}
Equation \eqref{eq:key} is the whole content of the theorem: the average
of a function of the stochastic variable is the expectation value of the
same function of the operator, taken on the reference state. Applying it
to $f(\ell)=\mel{x}{\Res(\{\ell\};E)}{y}$ for each variable in turn, and
using the fact that the $M_k$ act on distinct tensor factors and therefore
commute with one another, the average over the product distribution
\eqref{eq:avdef} becomes the matrix element \eqref{eq:theorem}.
$\blacksquare$

Two remarks. Equation \eqref{eq:key} is proved for a scalar function $f$,
while it is applied to a matrix element of the resolvent, which is an
operator in $\Hil$; the extension is legitimate because each $M_k$ acts on
its own tensor factor of $\Phi$ and therefore commutes both with the other
$M_{k'}$ and with everything carrying indices in $\Hil$, so that the
substitution $\ell_k\to M_k$ may be performed inside every term of the
expansion of the resolvent in powers of the disorder. Second, one may
equally well work in the basis that diagonalises $M$, since
\begin{equation}
  \mel{v_0}{f(M)}{v_0}=
  \mel{v_0}{U^{\dagger}U f(M) U^{\dagger}U}{v_0}=
  \mel{x_0}{f(D)}{x_0},
  \label{eq:unitary}
\end{equation}
with $D=UMU^{\dagger}$ diagonal and $\ket{x_0}=U\ket{v_0}$.

\section{Displacement generators}
\label{app:gener}

The generators \eqref{eq:Mgen} are Hermitian, commute among themselves and
obey the closure relation \eqref{eq:closure}. Hermiticity is immediate
from the reality of $Y_{10}$ and from
$Y^{*}_{\ell m}=(-1)^{m}Y_{\ell-m}$, which relates $M^{+}$ and $M^{-}$.
Commutativity follows from \eqref{eq:Jode}: since
$\partial_{u_\alpha}\partial_{u_\beta}J=\partial_{u_\beta}
\partial_{u_\alpha}J$ and $J$ is invertible,
$[M^{\alpha},M^{\beta}]=0$. Closure follows from
$\sum_{\alpha}k_\alpha k_\alpha=k^{2}$ together with the completeness of
the spherical harmonics on the unit sphere,
\begin{equation}
  \sum_{L}Y^{*}_{L}(\hat r)Y_{L}(\hat r')
  =\delta(\varphi-\varphi')\,\delta(\cos\theta-\cos\theta') .
\end{equation}

One caveat is important in practice. Both \eqref{eq:Mcomm} and
\eqref{eq:closure} involve a sum over intermediate angular momenta, and
therefore hold exactly only in the complete basis. In a calculation
truncated at $\ell_{\max}$ the sum is cut, and the violation at the
boundary channel $\ell=\ell_{\max}$ is not small: it is $O(1)$, since the
missing term is of the same order as those retained. Away from the
boundary the identities are satisfied to machine precision. The practical
rule is therefore that $\ell_{\max}$ must exceed the physically relevant
channels by at least one, the outermost shell of the basis acting as a
buffer rather than as a contributing channel.

A second remark concerns the angular basis, since the two used in this
work are not the same. The generators \eqref{eq:Mgen}, the identity
\eqref{eq:cart2sph} and the commutator \eqref{eq:comm-explicit} are
written in the
basis of complex spherical harmonics, where the Gaunt integrals and the
Clebsch--Gordan algebra take their standard form, whereas the recursion of
Appendix~\ref{app:cf} is run in the basis of real harmonics, which is what
makes the augmented operator complex symmetric. The two are related by a
fixed unitary matrix $U$, block diagonal in $\ell$, and every object in
the correction transforms by the same congruence: $M^{\alpha}\to
UM^{\alpha}U^{\dagger}$, $\Pi^{ii}\to U\Pi^{ii}U^{\dagger}$, and a
$t$-matrix diagonal in $\ell$ is left invariant, $U t U^{\dagger}=t$,
since $U$ commutes with any multiple of the identity within an $\ell$
block. Hence
\begin{equation}
  D\;\longrightarrow\;U D\,U^{\dagger},
  \label{eq:Dcov}
\end{equation}
that is, $D$ is one and the same operator in the two bases, and only its
matrix elements are relabelled. The Cartesian index $\alpha$ plays no part
in this, being unrelated to the angular basis. One may therefore evaluate
the correction through \eqref{eq:Dsph} and \eqref{eq:comm-explicit}, where the angular algebra is
transparent, and transform the result, or work throughout in the real
basis; the choice is one of convenience, and only the recursion, which
requires $\mathcal{A}^{\mathsf T}=\mathcal{A}$, has to be carried out in
the real basis.

\section{Continued-fraction recursion}
\label{app:cf}

\subsection{The Lanczos scheme}

Given a Hermitian operator $M$ and a normalised starting state
$\ket{x_0}$, with $\ket{x_{-1}}=0$, one constructs the chain
\begin{equation}
  \ket{x_{n+1}}=M\ket{x_n}-a_n\ket{x_n}-b_n^{2}\ket{x_{n-1}},
\end{equation}
with the coefficients fixed by requiring each new vector to be orthogonal
to the previous ones,
\begin{equation}
  a_n=\frac{\mel{x_n}{M}{x_n}}{\braket{x_n}{x_n}},
  \qquad
  b_n^{2}=\frac{\mel{x_{n-1}}{M}{x_n}}{\braket{x_{n-1}}{x_{n-1}}} .
\end{equation}
Suitably normalised, the $\ket{x_n}$ form a basis in which $M$ is
tridiagonal, and the diagonal element of the resolvent is the
$J$-fraction \eqref{eq:jfraction} with those coefficients. The scheme
generalises to non-Hermitian operators through a two-sided recursion with
distinct left and right vectors~\cite{Haydock,GrossoPastori}.

For the augmented operator a stronger statement is available, and it is
worth making because it determines the choice of angular basis. In the
basis of \emph{real} spherical harmonics the multiple-scattering matrix is
complex symmetric rather than Hermitian, $\mathcal{A}^{\mathsf T}
=\mathcal{A}$, and so is its promotion, since the disorder blocks
$M^{(i)}$ are real symmetric and the off-diagonal part is
$G\otimes\id_\Phi$. A complex symmetric operator admits a single
three-term recursion, with $\braket{x_n}{x_n}$ read as the bilinear rather
than the sesquilinear form; the two-sided scheme is unnecessary. In the
complex basis this is false: there the matrix satisfies
$\mathcal{A}^{\mathsf T}=\mathcal{J}\mathcal{A}\mathcal{J}$ with
$\mathcal{J}$ the conjugation $(\ell,m)\to(-1)^{m}(\ell,-m)$, and a
three-term recursion silently produces a projection that is not
tridiagonal, and a continued fraction that converges to the wrong value.
Real harmonics are in any case the convention of
Refs.~\cite{NatoliPRB90,TysonPRB92}. The displacement generators of
Sec.~\ref{sec:displ} are, on the other hand, written in the complex basis,
where the angular algebra is standard; that this entails no ambiguity in
the repeated-site correction is shown at the end of
Appendix~\ref{app:gener}.

The usual caveat of the complex symmetric algorithm applies: the bilinear
form is not positive definite, so an isotropic vector with
$\braket{x_n}{x_n}=0$ can in principle break the recursion. The standard
remedy is a look-ahead step. In the tests reported in
Sec.~\ref{sec:validation} the situation did not occur.

Two remarks connect this with what follows. The operator to which the
symmetric three-term recursion applies is $\mathcal{A}=\mathcal{T}_a^{-1}+G$
of \eqref{eq:Aaug}, whose inverse gives the configuration average directly
through \eqref{eq:tauaug}, and it is on $\mathcal{A}$ that the calculation
is actually carried out. The two subsections below instead work with the
arrangement $1+GT$ of Filipponi~\cite{Filipponi1991}, which is the natural
one for exhibiting the multiple-scattering series level by level, the two
being related by
\begin{equation}
  \tau^{00}=t^{0}\bigl[(1+GT)^{-1}\bigr]^{00},
  \label{eq:tauGT}
\end{equation}
since $T$ is block diagonal in the site index. That operator is \emph{not}
symmetric, even in the real basis: there $G$ and $T$ are separately
symmetric, so that $(1+GT)^{\mathsf T}=1+TG$, which differs from $1+GT$
whenever $[G,T]\neq0$. The recursion on $1+GT$ therefore does require
distinct left and right vectors, and is written that way below. The choice
between the two arrangements is one of convenience: $1+GT$ makes the path
structure visible, $\mathcal{A}$ makes the recursion symmetric.

\subsection{The recursion for the ordered cluster}

We first recall how the recursion generates the multiple-scattering
series, following Filipponi~\cite{Filipponi1991}. The object to be
computed is $[(\id+GT)^{-1}]^{00}_{L_0L_0}$, the cluster consisting as in
the body of the absorber $i=0$ and its $N$ neighbours, with
\begin{equation}
  T=\mathrm{diag}\{t^{0},t^{1},t^{2},\dots\},
  \qquad
  G=\begin{pmatrix}0&G^{01}&G^{02}&\cdots\\
  G^{10}&0&G^{12}&\cdots\\
  G^{20}&G^{21}&0&\cdots\\
  \vdots&&&\ddots\end{pmatrix}.
\end{equation}
The states are grouped in blocks of site index, each block carrying the
angular momenta. In this subsection the starting vector is numbered
$\ket{1}$; consequently $a_1$ corresponds to $a_0$ in the convention of
Eq.~\eqref{eq:jfraction}. Taking $\ket{1}=(1,0,0,\dots)^{\mathsf T}$ as
the starting state, and noting that $G$ has no diagonal blocks,
\begin{equation}
  a_1=\mel{1}{\id+GT}{1}=1,
\end{equation}
while
\begin{equation}
  b_1\ket{2}=\bigl(\id+GT\bigr)\ket{1}-a_1\ket{1}
  =\bigl(0,\;(G^{10}t^{0})_{LL_0},\;\dots,\;
  (G^{J0}t^{0})_{LL_0},\;\dots\bigr)^{\mathsf T},
\end{equation}
and correspondingly for the left vector, which is built by the same steps
applied to the transpose, as the two-sided scheme requires here, $1+GT$
not being symmetric, so that
\begin{equation}
  b_1^{2}=\sum_{J=1}^{N}
  \bigl(G^{0J}t^{J}G^{J0}t^{0}\bigr)_{L_0L_0},
  \label{eq:b1}
\end{equation}
which is the single-scattering term. Continuing the chain generates the
multiple-scattering series, the level $n$ collecting the paths of order
$n$.

For a cluster of two atoms and $\ell,\ell'=0,1$ everything can be written
out. With
\begin{equation}
  T=\begin{pmatrix}t^{0}_{00}&&&\\&t^{0}_{11}&&\\&&t^{1}_{00}&\\
  &&&t^{1}_{11}\end{pmatrix},
  \qquad
  GT=\begin{pmatrix}0&0&G^{01}_{00}t^{1}_{00}&G^{01}_{01}t^{1}_{11}\\
  0&0&G^{01}_{10}t^{1}_{00}&G^{01}_{11}t^{1}_{11}\\
  G^{10}_{00}t^{0}_{00}&G^{10}_{01}t^{0}_{11}&0&0\\
  G^{10}_{10}t^{0}_{00}&G^{10}_{11}t^{0}_{11}&0&0\end{pmatrix},
\end{equation}
and $\ket{1}=(1,0,0,0)^{\mathsf T}$ selecting $L_0=0$, one finds
\begin{equation}
  b_1^{2}=G^{01}_{00}t^{1}_{00}G^{10}_{00}t^{0}_{00}
         +G^{01}_{01}t^{1}_{11}G^{10}_{10}t^{0}_{00},
\end{equation}
the first-shell formula for two atoms.

\subsection{The recursion in the augmented space}

The same construction applies verbatim to the augmented operator, the only
change being the starting state, which now carries the reference state of
the configuration space:
\begin{equation}
  \ket{1F}=\begin{pmatrix}1\\0\\\vdots\end{pmatrix}
  \otimes\ket{F}\equiv\ket{x_0},
\end{equation}
where $\ket{F}$ is the reference state of whichever configuration space is
at issue, $\bigotimes_{i}\ket{f_0^{(i)}}$ for substitutional disorder and
$\bigotimes_{i\alpha}\ket{f_0^{(i\alpha)}}$ for thermal disorder. We write
out the thermal case, which is the less obvious of the two, and keep the
displacements on the propagator rather than transferring them to the
scatterers as in \eqref{eq:bridge}: $\mathcal{G}(u)$ below denotes the
promoted propagator $\mathcal{J}(\vec R)\,G\,\mathcal{J}(-\vec R)$, with
$G$ evaluated between the equilibrium positions. Since it still has no
diagonal blocks,
\begin{equation}
  a_1=\mel{x_0}{\id+\mathcal{G}(u)\mathcal{T}_a}{x_0}=1,
\end{equation}
exactly as before, and
\begin{equation}
  b_1\ket{x_1}=\mathcal{G}(u)\mathcal{T}_a\ket{x_0}
  =\bigl(0,\dots,(\mathcal{G}^{J0}t^{0})_{LL_0},\dots\bigr)^{\mathsf T}
  \otimes\ket{F} .
\end{equation}
Because the disorder variables of distinct sites are uncorrelated, each
$M^{(i)}$ acts only on its own factor, and the second-level coefficient is
\begin{equation}
  b_1^{2}=\sum_{J=1}^{N}
  \Bigl(G^{0J}\,\bar T^{J}\,G^{J0}\,t^{0}\Bigr)_{L_0L_0},
  \qquad
  \bar T^{J}=\mel{f_0^{(J)}}{J(-u_J)\,t^{J}\,J(u_J)}{f_0^{(J)}},
\end{equation}
with $\bar T^{J}$ the averaged $t$-matrix of Sec.~\ref{sec:DW}, the outer
factors $J(\pm u_0)$ having been dropped by the argument of
Sec.~\ref{sec:thermaug}. Comparison with
\eqref{eq:b1} shows that the recursion in the augmented space is the
recursion of the ordered cluster with every $t$-matrix replaced by its
average, as it must be, at this level, since each site is visited once.
The corrections of Sec.~\ref{sec:Dcorr} appear at the levels that generate
paths revisiting a site.

One feature of the recursion is worth recording, because it is what makes
an implementation practicable. For binary substitutional disorder the
augmented vectors are strictly local in the configuration space. Each term
of $\mathcal{A}$ acts on at most one factor of $\Phi$, and the propagator
acts on none, so that raising the occupation of a factor requires first
reaching the corresponding site: two applications of $\mathcal{A}$ per
factor, the absorber contributing none since it carries no disorder
operator. By induction the Krylov vector at level $m$ has strictly
vanishing, not merely small, amplitude on every configuration differing
from $\ket{F}$ in more than $\lfloor m/2\rfloor$ sites. The recursion may
therefore be run, with no approximation whatsoever, in the subspace
spanned by those configurations alone, whose dimension is
$(N+1)(\ell_{\max}+1)^{2}\sum_{j\le\lfloor m/2\rfloor}\binom{N}{j}$
instead of $(N+1)(\ell_{\max}+1)^{2}2^{N}$. Since the number of levels
required for convergence is governed by the scattering order rather than
by the size of the space, the sum is over a fixed number of terms and the
dimension is polynomial in $N$.

Two qualifications should be attached to that count. For a local factor of
dimension $\mu$ the same argument gives the corresponding upper bound with a
factor $(\mu-1)^{j}$, since a site excited away from its reference state
may sit in any of $\mu-1$ configurations. For continuous disorder the
local factor is infinite-dimensional and $\mathcal{J}(R)$ is dense on it,
so that a single application may populate many levels of the same site;
there the local Jacobi space must additionally be truncated, although the
restriction on the \emph{number of sites} reached at a given recursion
level remains valid as stated.

\section{Born limit for a square well}
\label{app:born}

To gauge the size of the differences $\Delta_\ell=t_\ell-t_{\ell+1}$ that
control the correction \eqref{eq:comm-explicit}, it is enough to evaluate
them in the Born approximation. In the normalisation
\eqref{eq:tmat} the Born $t$-matrix reads
\begin{equation}
  t_\ell\simeq\delta_\ell\simeq
  -\frac{2m}{\hbar^{2}}\,k\int_0^{\infty}
  r^{2}\bigl[j_\ell(kr)\bigr]^{2}V(r)\,\dd r,
  \label{eq:born}
\end{equation}
the factor $k$ being what distinguishes the dimensionless convention from
the one in which $t_\ell=k^{-1}e^{\ii\delta}\sin\delta$. The prefactor
$2m/\hbar^{2}$ equals unity in Rydberg units and two in the Hartree atomic
units used for the well depths of Sec.~\ref{sec:validation}; we retain it
explicitly below. For a square well
of depth $V_0$ and radius $R_0$,
\begin{equation}
  V(r)=\begin{cases}-V_0, & r\le R_0,\\ 0,& r>R_0,\end{cases}
  \qquad X\equiv kR_0,
\end{equation}
the two lowest channels are
\begin{align}
  t_0&=\frac{2m}{\hbar^{2}}\,\frac{V_0}{2k^{2}}
       \Bigl[X-\sin X\cos X\Bigr],\\
  t_1&=\frac{2m}{\hbar^{2}}\,\frac{V_0}{2k^{2}}
       \Bigl[X+\sin X\cos X-\frac{2\sin^{2}X}{X}\Bigr],
\end{align}
the two minus signs, that of \eqref{eq:born} and that of the attractive
well, having cancelled. In the difference the terms linear in $X$ cancel
while the two $\sin X\cos X$ add:
\begin{equation}
    t_0-t_1=\frac{2m}{\hbar^{2}}\,\frac{V_0}{k^{2}}
  \left[\frac{\sin^{2}X}{X}-\sin X\cos X\right]
  \label{eq:t0t1}
\end{equation}
For a weak, small scatterer the bracket behaves as $X^{3}/3$, giving
$t_0-t_1\simeq\tfrac13\,(2m/\hbar^{2})\,V_0kR_0^{3}$: the correction
switches off as the
scatterer becomes pointlike, as it must, and grows linearly with $k$ at
fixed geometry. At large $X$ the bracket remains oscillatory and of order
unity; the prefactor therefore makes $t_0-t_1$ decay algebraically as
$k^{-2}$ in this Born model, so that the differences between consecutive
channels survive well into the EXAFS region, and with them the
repeated-site correction. The overall sign of \eqref{eq:t0t1} is
immaterial in the quadratic correction \eqref{eq:Ddiag}, although the
complex phase of $\Delta_\ell$ is retained in the general case.

\section{Ternary alloys}
\label{app:ternary}

Nothing in the construction is tied to two components. We record here the
ternary case, both because it is the first one in which the projector
identity $M^{2}=M$ ceases to hold and because the number of distinct
configurations, $3^{N}$ instead of $2^{N}$, puts an enumeration out of
reach at cluster sizes where the recursion is unaffected: nothing in
Appendix~\ref{app:cf} refers to the number of components, and the
confinement argument given there holds unchanged.

Let the three species $A$, $B$, $C$ occur with concentrations $x$, $y$,
$z$, and let the configuration variable of a site take the three values
$E_i=1,0,-1$ with
\begin{equation}
  p(E_i)=x\,\delta(E_i-1)+y\,\delta(E_i)+z\,\delta(E_i+1) .
  \label{eq:pternary}
\end{equation}
The inverse $c^{i}=(t^{i})^{-1}$ of the scatterer must reduce to $c^{A}$,
$c^{B}$, $c^{C}$ at the three values. The unique quadratic interpolation
through those three points is
\begin{equation}
  c^{i}
  =c^{B}+E_i\bigl(c^{A}-c^{B}\bigr)
   -\frac{E_i(1-E_i)}{2}\bigl(c^{C}-2c^{B}+c^{A}\bigr)
  \equiv c^{B}+E_i\,\Delta\alpha-\frac{E_i(1-E_i)}{2}\,\Delta\beta,
  \label{eq:ternenc}
\end{equation}
with $\Delta\alpha=c^{A}-c^{B}$ the first difference and
$\Delta\beta=c^{C}-2c^{B}+c^{A}$ the second. The encoding is exact, not an
approximation: it is Lagrange interpolation on three nodes. In the binary
limit $z=0$ the variable takes only the values $1$ and $0$, the quadratic
term $E_i(1-E_i)$ vanishes identically, and \eqref{eq:ternenc} collapses
to the affine promotion $c^{i}=c^{B}+\Delta c\,n_i$ of
Sec.~\ref{sec:promotion}. The curvature term survives precisely because a
third species makes $c$ a genuinely quadratic function of the
configuration variable.

The moments of \eqref{eq:pternary} are $\mu_0=1$ and, for $n\ge1$,
\begin{equation}
  \mu_n=x+z\,(-1)^{n}
  =\begin{cases} s, & n \text{ even},\\ d, & n \text{ odd},\end{cases}
  \qquad s\equiv x+z,\quad d\equiv x-z,
\end{equation}
so that the Hankel determinants \eqref{eq:hankeldef} are
\begin{equation}
  D_0=1,\qquad
  D_1=s-d^{2},\qquad
  D_2=\bigl(s^{2}-d^{2}\bigr)(1-s)=4xyz,\qquad
  D_3=D_4=\dots=0 .
  \label{eq:ternhankel}
\end{equation}
The vanishing of $D_3$ and of all that follow is the structural fact: a
measure supported on three points has vanishing Hankel determinants from
order three onwards, exactly as the binary measure has them vanishing from
order two. The recursion therefore terminates, and the configuration
factor of each site is a space of dimension three, $\dim\phi^{(i)}=3$ and
$\dim\Phi=3^{N}$.

Reading the coefficients off \eqref{eq:hankel}, with $R_0=\mu_1=d$,
$R_1=\mu_3-\mu_1\mu_2=d\,y$ and $R_2=0$, the last because two rows of the
determinant coincide, gives $a_0=d$, $a_1=\kappa-d$, $a_2=-\kappa$, and
the disorder operator of a single site is the $3\times3$ Jacobi matrix
\begin{equation}
    M=\begin{pmatrix}
    d & b_1 & 0\\
    b_1 & \kappa-d & b_2\\
    0 & b_2 & -\kappa
  \end{pmatrix},
  \qquad
  \kappa=\frac{d\,y}{s-d^{2}},
  \qquad
  b_1^{2}=s-d^{2},
  \qquad
  b_2^{2}=\frac{4xyz}{\bigl(s-d^{2}\bigr)^{2}}
  \label{eq:Mternary}
\end{equation}
its trace vanishing, as it must, because the eigenvalues are the support
$\{1,0,-1\}$ of \eqref{eq:pternary}. The characteristic polynomial is
$\lambda^{3}-\lambda$, so that in place of the binary projector identity
one has
\begin{equation}
  M^{3}=M,
  \label{eq:M3}
\end{equation}
and $M$ is no longer idempotent, which is the algebraic counterpart of
the surviving quadratic term in \eqref{eq:ternenc}. For the symmetric
composition $x=z$ the diagonal empties, $d=\kappa=0$, and the matrix
reduces to the bare chain $b_1=\sqrt{s}$, $b_2=\sqrt{y}$ with
$b_1^{2}+b_2^{2}=1$.

The spectral argument of Sec.~\ref{sec:promotion} carries over unchanged.
The three eigenprojectors $P_A$, $P_B$, $P_C$ of $M$, associated with the
eigenvalues $1$, $0$, $-1$, are orthogonal and complete, so that the
promoted scatterer is again a spectral decomposition,
\begin{equation}
  \mathcal{T}^{i}=t_A^{i}\otimes P_A^{(i)}+t_B^{i}\otimes P_B^{(i)}
  +t_C^{i}\otimes P_C^{(i)},
\end{equation}
and $f(\mathcal{T}^{i})=\sum_{s}f(t^{i}_{s})\otimes P_s^{(i)}$ for any
function defined on the three blocks. Promoting $t$ and promoting $t^{-1}$
remain simultaneously exact, for non-commuting angular-momentum matrices
as before, and \eqref{eq:tauaug} holds verbatim with
$\ket{F}=\bigotimes_i\ket{f_0^{(i)}}$ and
$\ket{f_0}=(1,0,0)^{\mathsf T}$.
Everything else in the paper is untouched.


\end{document}